\documentclass[a4paper, 11pt]{elsarticle} 

\makeatletter
\def\ps@pprintTitle{%
	\let\@oddhead\@empty
	\let\@evenhead\@empty
	\let\@oddfoot\@empty
	\let\@evenfoot\@oddfoot
}
\makeatother
\usepackage{lineno,hyperref,indentfirst,caption,float,amsmath,tabularx,algorithm,setspace,longtable,multirow, graphicx, booktabs, array, caption,enumitem,cleveref,color,soul,lscape, appendix, amssymb}

\allowdisplaybreaks
 
\usepackage[margin=1in]{geometry}
\usepackage{url}
\usepackage{algorithm,algpseudocode}
\newcolumntype{P}[1]{>{\centering\arraybackslash}p{#1}}
\newcolumntype{M}[1]{>{\centering\arraybackslash}m{#1}}
\makeatletter
\def\BState{\State\hskip-\ALG@thistlm}
\makeatother

\journal{Transportation Research Part E}
\biboptions{numbers,sort&compress}

\usepackage{hanging}
\usepackage{tikz}

\usetikzlibrary{calc,positioning,shapes.geometric, arrows}
\tikzstyle{arrow} = [thick,->,>=stealth]

\tikzset{
	every node/.style={draw=black, text centered,text width=15cm},
}

\usepackage{titlesec}
\titleformat{\subsection}
{\bfseries\itshape}{\thesubsection}{1em}{}

\begin{document}
\begin{center}	
\vspace{1cm}

\textbf{\Large Air Taxi Service for Urban Mobility: A Critical Review of Recent Developments, Future Challenges, and Opportunities}

\vspace{1cm}
\textbf{\large{Suchithra Rajendran and Sharan Srinivas}}
\setstretch{1}
\text{Department of Industrial and Manufacturing Systems Engineering, College of Engineering,}
\text{Department of Marketing, Trulaske College of Business}
\text{University of Missouri, Columbia, MO 65211, USA}\\
\end{center}	
\vspace{0.5 cm}

\setstretch{1}
\noindent This is a pre-print version of the manuscript accepted in Transportation Research Part E: Logistics and Transportation Review (\url{https://doi.org/10.1016/j.tre.2020.102090})

\vspace{0.5 cm}
\noindent  The article should be cited as follows: Rajendran, S., \& Srinivas, S. (2020). Air taxi service for urban mobility: a critical review of recent developments, future challenges, and opportunities. Transportation research part E: logistics and transportation review, 143, 102090.

\begin{center}
	\setstretch{1}	
\vspace{15cm}
\color{darkgray}© 2020. This manuscript version is made available under the CC-BY-NC-ND 4.0 license \url{http://creativecommons.org/licenses/by-nc-nd/4.0/}

\end{center}

\begin{frontmatter}
\title{Air Taxi Service for Urban Mobility: A Critical Review of Recent Developments, Future Challenges, and Opportunities}

\author{Suchithra Rajendran\corref{cor1}}
\ead{RajendranS@missouri.edu}

\author{Sharan Srinivas\corref{cor2}}

\cortext[cor1]{Corresponding author}

\address{\scriptsize Department of Industrial and Manufacturing Systems Engineering, University of Missouri, Columbia, MO 65211, USA \\ Department of Marketing, University of Missouri, Columbia, MO 65211, USA  \vspace{-12mm}}

\begin{abstract}
	
Expected to operate in the imminent future, air taxi service (ATS) is an aerial on-demand transport for a single passenger or a small group of riders, which seeks to transform the method of everyday commute. This uncharted territory in the emerging transportation world is anticipated to enable consumers bypass traffic congestion in urban road networks. By adopting an electric vertical takeoff and landing concept (eVTOL), air taxis could be operational from skyports retrofitted on building rooftops, thus gaining advantage from an implementation standpoint. Motivated by the potential impact of ATS, this study provides a review of air taxi systems and associated operations. We first discuss the current developments in the ATS (demand prediction, air taxi network design, and vehicle configuration). Next, we anticipate potential future challenges of ATS from an operations management perspective, and review the existing literature that could be leveraged to tackle these problems (ride-matching, pricing strategies, vehicle maintenance scheduling, and pilot training and recruitment). Finally, we detail future research opportunities in the air taxi domain. 
\medskip
\end{abstract}

\end{frontmatter}


\section{Introduction}
\subsection{Background}

Traffic congestion has become a ubiquitous problem in urban areas due to various reasons, such as high population density, increase in privately-owned vehicles, rise in the number of commuters traveling from low-density areas where public transits do not serve passengers effectively, and average household income inflation (Downs, 2005; Koźlak and Wach, 2008). In metropolitan cities, such as Los Angeles and New York, an average commuter spends over 90 minutes in traffic resulting in an increase in stress and anxiety (Kawabata and Shen, 2006; Inrix, 2018). Such congestions also lead to 330 grams per mile of $\text{CO}_2$ emissions into the atmosphere (Barth and Boriboonsomsin, 2009). Besides having a detrimental impact on the health of the commuters and the environment, it also contributes to economic loss. For instance, Manhattan experiences an annual loss of \$20 billion due to traffic congestion, with the excess fuel and vehicle operating cost accounting for nearly 13\% of the total loss (Partnership for New York City, 2020). Similar economic impacts are also experienced by other congested cities across the globe, such as Bangalore, London, Paris, and Tokyo (Downs, 2005; Ikeuchi et al., 2019). Therefore, it is imperative to explore beyond ground transportation modes for facilitating faster everyday commute for passengers in urban areas and alleviating road congestion. 

Several logistics companies and aviation agencies are striving to leverage urban air mobility (UAM) using flying taxi services, an aviation ride-hailing concept, that is expected to launch in the forthcoming years (Garrow et al., 2018; Rajendran and Zack, 2019; Swadesir and Bil, 2019). While aircraft makers such as, Airbus, Lilium, and Kitty Hawk, have been actively involved in  air taxi manufacturing over the past few years (O’Hear, 2017; Hawkins, 2018; Warwick, 2018), companies like Bell, Embraer, Hyundai, Rolls-Royce, and Toyota have recently ventured into the air taxi market (Hawkins, 2020; Hornyak, 2020; Sampson, 2020). Besides Uber, which estimates the launch of its air taxis (called Uber Elevate) in 2023, other companies such as Zephyr Airworks and Airbus are also currently taking measures to conduct tests with their electric aviation taxis, namely Cora and Airbus-Vahana respectively, in countries across the world, such as the USA, Japan, Singapore, New Zealand, France, and India (Pointon, 2018; Rajendran and Zack, 2019). Through the use of proposed fully electric vertical takeoff and landing (eVTOL) vehicles, air taxi service (ATS) could offer a faster and reliable mode of transportation. Besides, the usage of electric air taxis is anticipated to be more energy-efficient, safer, and quieter than any modern helicopter (Polaczyk et al., 2019; Pradeep and Wei, 2019). Table \ref{tab:Tab1} summarizes the list of terminologies used in ATS.

\begin{table}[H]
  \centering
  \caption{Common terminologies in air taxi transportation}.
  \scalebox{0.8}
   {
   	    	\setlength{\tabcolsep}{10pt} 
   	\renewcommand{\arraystretch}{1.5} 
   	\begin{tabular}{m{15em}m{35em}}
    \toprule
    \centering\textbf{Term} & {\centering\textbf{Definition}} \\
    \midrule
    Air Taxi & {Small electric aircraft that are expected to transport on-demand passengers in metropolitan cities (Saurin and Junior, 2012; Antcliff et al., 2016; Johnson et al., 2018), with an average capacity of four.} \\
    eVTOL & Electric vertical takeoff and landing (eVTOL) technology is similar to that of a helicopter, allowing air taxis to lift and land vertically or maneuver at precise vertical angles between the buildings and other obstacles within a large city (Johnson et al., 2018; Ng and Datta, 2019).\\
    Vertiport & A large centralized hub with facilities for customer pick-up or drop-off, charging, maintenance, and several vehicle docking stations (Mueller et al., 2017; Dunn, 2018; Rothfeld et al., 2018).  \\
    Vertistop & A site dedicated only for customer pick-up or drop-off, with no support infrastructure, such as charging stations (Smith, 1994; Patnoe, 2018; and Hasan, 2019)\\
    Station/Infrastructure/Skyport & A common term used for referring to both vertiport and vertistop (Rajendran and Shulman, 2020) \\
    \bottomrule
    \end{tabular}
}%
  \label{tab:Tab1}%
\end{table}%

Due to the rapid advancement in urban air mobility (UAM), several recent research has investigated the design concepts and implementation issues (Vascik and Hansman, 2018; Rajendran and Pagel, 2020). Recently, Straubinger et al. (2020) provided a review of UAM regulations, aircraft design, requirements, certifications, policies, and traffic management. Our study not only focusses on the recent design developments, but also intensely surveys prior works on the qualitative and quantitative approaches for infrastructure and demand estimation. Besides, we focus on challenges and future research opportunities from an operations management perspective, while their work is geared towards implementation issues, such as policies and certifications.

\begin{figure}[H]
	\centering
	\captionsetup{justification=centering}
	\includegraphics[width=0.75\linewidth]{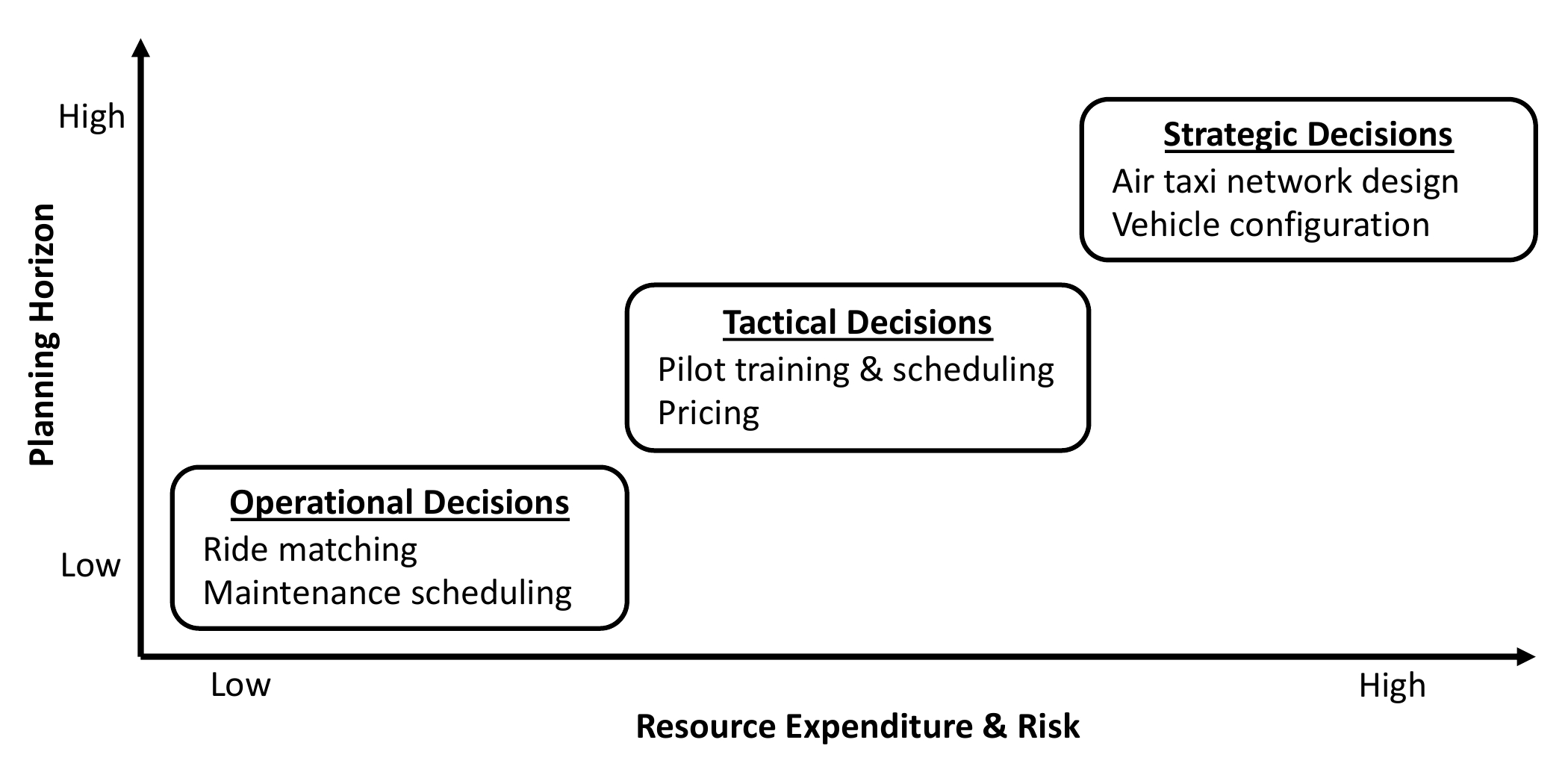}
	\caption{Categorization of air taxi operations into three management decision levels}
	\label{fig:Fig1}
\end{figure}

The introduction of these ATS would be associated with several implementation and operational challenges, which can be categorized into three management decision levels - strategic, tactical, and operational. Strategic decisions have a long-term impact on ATS operations and are characterized by high capital expenditures and risk. Air taxi network decisions, such as determining the number and location of vertiports/vertistops, are typical strategic decisions. On the other hand, choices that have a medium-term impact on the aviation company’s resources, such as pilot recruitment, are referred to as tactical decisions. Operational decisions are short-term and have relatively less expenditure and risk, such as real-time routing of air taxis. Moreover, these three decisions could also be interrelated. For instance, the number and location of vertistops affect the pricing and routing decisions. Thus, it is necessary to optimize the decisions at all levels to efficiently route and coordinate the movement of air taxis throughout the network. Figure \ref{fig:Fig1} shows the categorization of air taxi design and operations under the three decision levels.  

\subsection{Overview of Urban Air Taxi Service
}
An ATS shares the characteristics of common transportation modes, such as regular taxis and subways, as shown in Table \ref{tab:Tab2}. The experience of requesting an ATS would be similar to the process of booking a standard on-demand mobility (ODM) service. Typically, a registered customer will enter the pickup and drop-off locations using a ridesharing application. Depending on the trip information, the platform will estimate the fare amount and travel time for all applicable transportation modes, such as air and regular taxis. If an ATS is feasible, then a customer may avail of the service depending on several attributes like willingness-to-fly, trip cost, and transit-time sensitivity (Rajendran and Shulman, 2020). 

\begin{figure}[!]
	\centering
	\captionsetup{justification=centering}
	\includegraphics[width=0.75\linewidth]{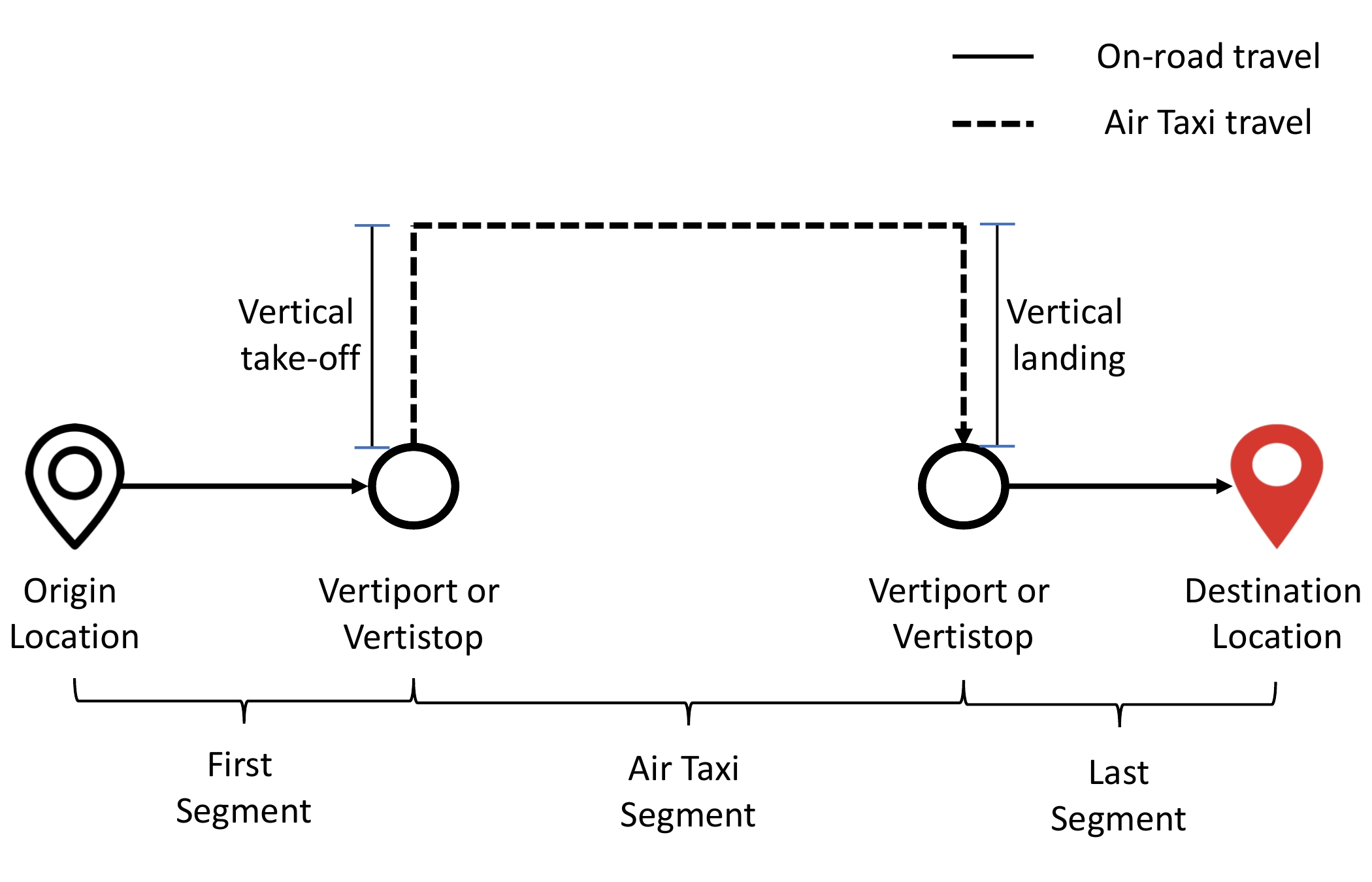}
	\caption{Schematic of a typical air taxi trip}
	\label{fig:Fig2}
\end{figure}

An ATS is anticipated to have multiple segments, as shown in Figure \ref{fig:Fig2}. The first segment of the trip (commute from the pickup location to a vertiport/vertistop) is facilitated by the ridesharing platform via on-road car trips. Subsequently, the longest segment of the trip is covered by an air taxi, where the passenger is transported from the origin skyport to the destination station. Finally, the last mile of the trip (i.e., travel from destination skyport to the drop-off location) is again complemented by regular taxis. However, if the first or last segment of the trip is in close proximity to the designated skyport, then the customer may be directed to walk.

\begin{table}[htbp]
	\centering
	\caption{Relative comparison of various urban transport services based on their key characteristics}
	\scalebox{0.8}{ \begin{tabular}{lccc}
			\toprule
			\textbf{Characteristics} & \textbf{Regular Taxi} & \textbf{Subway and Buses} & \textbf{Air Taxi} \\
			\midrule
			On-Demand Availability & Yes   & No    & Yes \\
			Vehicle Routing & Real-time & Pre-defined & Real-time \\
			Number of Segments per Trip & Single & Single/Multiple & Single/Multiple \\
			Vehicle Capacity & Low   & High  & Low \\
			Trip Distance & Short, medium or long & Short, medium or long & Long or very long \\
			Ride Fare & Medium & Low   & High \\
			Vehicle Travel Speed & Low   & Moderate & High \\
			Trip Duration Uncertainty & High  & Low   & Low \\
			Ridesharing & Both shared and non-shared & Shared & Both shared and non-shared \\
			\bottomrule
	\end{tabular}}%
	\label{tab:Tab2}%
\end{table}%

\subsection{Contribution and Organization}
The aim of this paper is to provide the current state-of-the-art developments and future challenges in ATS. Towards this objective, we review both recent work in air taxi design and operations, as well as other relevant studies that are applicable to UAM. The contributions of this paper are three-fold. To the best of our knowledge, our article is the first to survey the recent research on air taxi systems predominantly from an operations management perspective. Second, potential unaddressed challenges associated with air taxi operations are identified, and insights to solve them are provided by reviewing research on other similar applications. Finally, several opportunities for future research are established to improve the operational efficiency of ATS. 

This paper is organized as follows. Section 2 summarizes the current developments in air taxi design and operations. The articles pertaining to the challenges associated with ATS, such as ride-matching, pricing strategies, vehicle maintenance scheduling, and pilot training and recruitment, are surveyed in Section 3. The opportunities for future research directions in the air taxi domain are reviewed in Section 4, while the concluding remarks are provided in Section 5.

\section{Current Developments}
This section details the current state-of-the-art developments in the air taxi system. We review both the past and recent work in the field of air taxi demand prediction, network design, and vehicle configuration.

\subsection{Demand Prediction for Air Taxis}
Akin to other ODM services, air taxi operations could endure inefficiencies due to the uncertainty associated with customer demand, which is geographically distributed and time-varying (Davis et al., 2018; Luo et al., 2020). The air taxi network includes a set of vertiports and vertistops,$\mathcal{V}=\{v_1,v_2,v_3…,v_S\}$, to facilitate passenger pickup and drop-off. Each skyport,  $s \in \mathcal{V}$, receives $\mathcal{D}_{st}$ customer requests during the time interval t. Besides, each customer request $c_j$ is defined as a tuple $(\phi_j^P,\phi_j^D,\lambda_j^P,\lambda_j^D)$, where $(\phi_j^P, \lambda_j^P)$ pair represents the pick-up latitude and longitude, while $(\phi_j^D, \lambda_j^D)$ is the coordinate of drop-off vertistop. If the pickup requests are not anticipated in advance, then it would increase the likelihood of having an imbalanced fleet of air taxis at the vertiports and vertistops - an oversupply of vehicles at some locations and insufficient fleet capacity at the remaining sites. This would lead to severe repercussions, such as high customer wait time and ineffective fleet utilization, which, in turn, adversely affects customer satisfaction and revenue. An accurate prediction of the demand for ODM services could facilitate better planning to overcome these challenges (Ma et al., 2014; Santi et al., 2014; Dandl et al., 2019). Moreover, demand estimation is critical to both fleet operators and transit authorities as their common goal is to reduce the number of empty air taxis crowding the airspace. 

Unfortunately, when forecasting the customer volume for on-demand air services is in its initial phase, a breadth of challenges is ever-present. First, there is a lack of historical data within the eVTOL sector, making it difficult to estimate the demand for ATS using analytical methods. Second, the cost of service is expected to be high in the near-term market resulting from capital requirements, which, in turn, could impact the customer base. Third, it is cumbersome to garner public perception of air taxis, based on factors, such as pre-boarding customer screening, eVTOL safety, and long-distance commuting preference. Finally, telecommuting has become an increasingly popular form of working, having increased yearly, thus estimating demand is extremely uncertain for these on-demand air services (NASA Mobility UAM Market Study, 2018).

There are two potential strategies to overcome the aforementioned inherent challenges and predict the customer requests for ATS - (i) adopting qualitative approaches, such as scientifically designed surveys and focus groups on obtaining the subjective opinion of potential customers and experts (e.g., Binder et al., 2018; Garrow et al., 2018; Garrow et al., 2019), (ii) approximating historical data from other ODM services (e.g., ridesharing service, traditional taxi service) and relying on quantitative methods/analytical models (Rajendran and Zack 2019). 

\subsubsection{Qualitative Approaches}
Currently, there is very limited research to estimate the potential demand for eVTOLs. Besides, all these studies use qualitative approaches due to the lack of historical data and seek to determine the market demand for the introductory phase of ATS (Binder et al., 2018; Garrow et al., 2018; Boddupalli, 2019; Garrow et al., 2019). One of the first works conducted focus groups to gauge the demand for on-demand air services (Garrow et al. 2018). The study estimated high-income households to be the early adopters of eVTOLs (due to their ability to spend more/higher value of time). In addition, the focus group revealed the top three market segments to attract customers as airport transfers, end-to-end city transfers, and daily commuters. Subsequently, some works have designed surveys to assess customers’ perceptions and attitudes that influence their particular demand for the use of air taxis (Binder et al., 2018; Boddupalli, 2019; Garrow et al., 2019). 

Binder et al. (2018) constructed a stated preference survey to estimate consumers’ willingness-to-pay for air taxis in densely populated areas in the United States. Garrow et al. (2019) developed a sampling strategy and comprehensive survey with about 100 questions, in which they considered the influence of several constructs (e.g., lifestyle, personality, perceptual, attitudinal, socio-demographic). These surveys were adapted by Boddupalli (2019) to obtain the preference of 2,500 commuters from five urban US cities and forecast their taste heterogeneity using a discrete choice model. Among the decision-making factors, travel time and cost were most influential, while income did not have a substantial impact. The main finding of the study is that the demand for air taxi is highly polarized with a sufficiently large persuadable customer base. 

Existing literature on commute mode choice could also provide insights on factors that could influence the demand for new transportation modes, such as air taxi (Schwanen and Mokhtarian, 2005; Zhou, 2012; Chee and Fernandez, 2013; Shen et al., 2016; Sivasubramaniyam et al., 2020). Several studies found economic factors, such as trip fare and individual’s income to be an important determinant of mode choice in an urban setting (Tyrinopoulos and Antoniou, 2013; Sivasubramaniyam et al., 2020)  Interestingly, the preference of higher-income individuals is conflicting as some preferred private transport, while others preferred transportation modes like on-demand taxis (Chee and Fernandez, 2013; Shen et al., 2016; Sivasubramaniyam et al., 2020). This could be due to various other factors identified in the literature, such as ease-of-use, parking spot availability, and crowding (Hess, 2001; Ye and Titheridge, 2017). Aside from economic indicators, age and gender also influenced the mode of transportation (Zhou, 2012; Chee and Fernandez, 2013).  For instance, it was observed that commuters between the ages of 35 and 44 had a higher preference for using a private car (Tyrinopoulos and Antoniou, 2013). In terms of gender, males were more likely to use public transportation (Chee \& Fernandez, 2013). 

The insights from these qualitative studies would be useful for making decisions in the launch phase of ATS, namely, infrastructure requirements (e.g., number of vertiports required, location of these vertiports), pricing strategies, and fleet procurement (e.g., number of air taxis to procure/manufacture). Moreover, these choices have a medium- to long-term impact on air taxi service providers as these high-risk decisions are likely to affect revenue and performance.

\subsubsection{Quantitative Approaches}
Given the representative data for $t$ time intervals, the analytical approach of air taxi demand prediction problem deals with the estimation of total customer requests, $F_{s,t+n}$, in each skyport $s$ during the next $n$ time intervals. Becker et al. (2018) was among the first to predict the long-term market demand for eVTOL urban mobility using a gravity model.  Based on eight socio-economic variables, they identified 26 potential high demand markets across the globe for implementing ATS. Since there is no existing data to model air taxi demand, Rajendran and Zack (2019) proposed a generic model to approximate the long- and short-term demand for ATS from traditional taxi trip data. If the air taxi has an average speed of $S^A$, then they estimated the air taxi travel time for trip $j$ ($\mathcal{T}_j^A) $ as the sum of inter-vertiport flight time and on-road travel time for passenger transport to and from the skyport, as shown in Equation (\ref{eq:1}), where hav($(\phi_j^P, \lambda_j^P)$,$(\phi_j^D, \lambda_j^D)$) is the Haversine distance between the stations, and $\mathcal{T}_j^R$  is traditional taxi travel time for trip $j$. Subsequently, they considered a traditional taxi trip to be eligible for an ATS if there is $\alpha\%$ savings in ride time (i.e., $E_j^A=1$ if $\mathcal{T}_j^A \leq (1-\alpha)\mathcal{T}_j^R$). Thus, the total expected demand for ATS is $\sum_j E_j^A$.

\begin{equation} \label{eq:1}
\begin{aligned}
\mathcal{T}_j^A = \frac{\text{hav}((\phi_j^P, \lambda_j^P),(\phi_j^D, \lambda_j^D))}{S^A} + f(\mathcal{T}_j^R) && \forall j\\
\end{aligned}
\end{equation}

However, once air taxi services are operational, it could adapt analytical approaches used by existing ODM services to predict customer demand. The ODM literature predominantly focuses on predicting the number of short-term customer requests (5 – 60 minutes) using time-series or machine learning approaches (Moreira-Matias et al., 2013; Jiang et al., 2019). Typically, most models are developed using GPS trace data as predictors (Davis et al. 2018; Zhao et al., 2019; Luo et al., 2020), while few studies combined data from multiple sources to increase prediction accuracy (Rodrigues et al., 2019). Besides, a common strategy is to adopt more than one error measure (Equations \ref{eq:2} – \ref{eq:5}) to evaluate the prediction models  (e.g., Ke et al., 2017; Xu et al., 2018; Jiang et al., 2019; Li and Wan, 2019). Table \ref{tab:Tab3} summarizes some prominent works on-demand prediction in ODM services. Such short-term estimates for ATS would be crucial for making effective decisions on routing and pricing (surge vs. regular). Thus, ATS demand prediction forms the basis for the three levels of decision making – strategic, tactical, and operational.

\begin{equation} \label{eq:2}
\begin{aligned}
\text{Mean Absolute Error} \; (MAE_s) = \frac{1}{T} \sum_{t=1}^{T} |\mathcal{F}_{s,t} - \mathcal{D}_{s,t}|  && \forall s\\
\end{aligned}
\end{equation}

\begin{equation} \label{eq:3}
\begin{aligned}
\text{Root Mean Squared Error} \; (RMSE_s) = \sqrt{\frac{1}{T} \sum_{t=1}^{T} (\mathcal{F}_{s,t} - \mathcal{D}_{s,t})^2}  && \forall s\\
\end{aligned}
\end{equation}

\begin{equation} \label{eq:4}
\begin{aligned}
\text{Mean Absolute Percentage Error} \; (MAPE_s) = \frac{1}{T} \sum_{t=1}^{T} \frac{|\mathcal{F}_{s,t} - \mathcal{D}_{s,t}|}{\mathcal{D}_{s,t}}  && \forall s\\
\end{aligned}
\end{equation}

\begin{equation} \label{eq:5}
\begin{aligned}
\text{Symmetric Mean Absolute Percentage Error} \; (sMAPE_s) = \frac{1}{T} \sum_{t=1}^{T} \frac{|\mathcal{F}_{s,t} - \mathcal{D}_{s,t}|}{\mathcal{D}_{s,t}+\mathcal{F}_{s,t}+1}  && \forall s\\
\end{aligned}
\end{equation}

\begin{table}[H]
  \centering
  \caption{Summary of relevant prominent works on demand prediction for ODM services}
    \scalebox{0.7}{
    	\setlength{\tabcolsep}{10pt} 
    	\renewcommand{\arraystretch}{1.5} 
    	\begin{tabular}{lp{5em}rp{5em}p{5em}}
    \toprule
    \multicolumn{1}{l}{\textbf{Author}} & \multicolumn{1}{p{7em}}{\centering\textbf{Predictive Time Horizon}} & \multicolumn{1}{p{7em}}{\centering \textbf{Data Source}} & \multicolumn{1}{p{10em}}{\centering\textbf{Analytical Model}} & \multicolumn{1}{p{7em}}{\centering \textbf{Evaluation Measures}} \\
    \midrule
    \multicolumn{1}{l}{Moreira-Matias et al. (2013)} & \multicolumn{1}{c}{30 minutes} & \multicolumn{1}{p{7em}}{\centering Real-time streaming data} & \multicolumn{1}{p{10em}}{ARIMA, time-varying Poisson, weighted time-varying Poisson} & \multicolumn{1}{p{7em}}{\centering sMAPE} \\
    \multicolumn{1}{l}{Davis, Raina, and Jagannathan (2016)} & \multicolumn{1}{c}{60 minutes} & \multicolumn{1}{p{7em}}{\centering GPS} & \multicolumn{1}{p{10em}}{LR, STL, TBATS, HW} & \multicolumn{1}{p{7em}}{\centering MAPE} \\
    \multicolumn{1}{l}{Ke et al. (2017)} & \multicolumn{1}{c}{60 minutes} & \multicolumn{1}{p{7em}}{\centering Network car order data} & \multicolumn{1}{p{10em}}{Fusion convolution LSTM, LSTM, MLP, MA, ARIMA, CNN} & \multicolumn{1}{p{7em}}{\centering RMSE, MAPE, MAE} \\
    \multicolumn{1}{l}{Davis et al. (2018)} & \multicolumn{1}{c}{15 mins, 60 mins} & \multicolumn{1}{c}{GPS} & \multicolumn{1}{p{10em}}{GT+STL, GT+TBATS, VT+STL, VT+TBATS} & \multicolumn{1}{p{7em}}{\centering sMAPE} \\
    \multicolumn{1}{l}{Xu et al. (2018)} & \multicolumn{1}{c}{5 - 60 minutes} & \multicolumn{1}{p{7em}}{\centering GPS} & \multicolumn{1}{p{10em}}{LSTM-RNN} & \multicolumn{1}{p{7em}}{\centering sMAPE, RMSE} \\
    \multicolumn{1}{l}{Jiang et al. (2019)} & \multicolumn{1}{c}{5 -15 mins} & \multicolumn{1}{p{7em}}{\centering Network car order data} & \multicolumn{1}{p{10em}}{LS-SVM} & \multicolumn{1}{p{7em}}{\centering MAE, RMSE, MAPE} \\
    \multicolumn{1}{l}{Rodrigues et al. (2019)} & \multicolumn{1}{c}{1 day} & \multicolumn{1}{p{7em}}{\centering Web and GPS} & \multicolumn{1}{p{10em}}{LSTM-RNN, DL-FC} & \multicolumn{1}{p{7em}}{\centering MAE, RMSE, MAPE} \\
    \multicolumn{1}{l}{Zhao et al. (2019)} & \multicolumn{1}{c}{4, 8, 12, 24 hours} & \multicolumn{1}{c}{GPS} & \multicolumn{1}{p{10em}}{LSTM-RNN, MLP, ARIMA, MC, LZW} & \multicolumn{1}{p{7em}}{\centering sMAPE} \\
    \multicolumn{1}{l}{Luo et al. (2020)} & \multicolumn{1}{c}{10 minutes} & \multicolumn{1}{c}{GPS} & \multicolumn{1}{p{10em}}{MTDL, STDL, SVM, kNN} & \multicolumn{1}{p{7em}}{\centering RMSE, MAPE, MAE} \\
    \midrule
    \multicolumn{5}{m{56em}}{Note: ARIMA: Auto Regressive Integrated Moving Average, DL-FC: Deep learning fully connected layers, GT: Geohash tessellation, HW: Holt Winters, kNN: k-nearest neighbors, LR: Linear regression, LS-SVM: Least squares support vector machines, LSTM-RNN: Long Short-term Memory Recurrent Neural Network, LZW: Lempel-Ziv-Welch, MA: Moving average, MTDL: Multi-task deep learning, CNN: Convolution neural network, MC: Markov chain, STL: Seasonal and Trend decomposition using Loess, STDL: Single-task deep learning, VT: Voronoi tessellation.} \\
    \end{tabular}}%
  \label{tab:Tab3}%
\end{table}%

\normalsize
\subsection{Air Taxi Network Design}
The establishment of intracity operational infrastructure is one of the significant barriers for air taxi implementation (Sun et al., 2018; Vascik and Hansman, 2018; Reiche et al., 2019; Rath et al., 2019; Tarafdar et al., 2019). Many factors contribute to a location being chosen as a strategic operating station, such as the demand density, existence of adequate area for safe departure/landing, availability of space for charging stations, and easy accessibility (Percoco, 2010; Duval, 2013; Yang and Notteboom, 2016; Bozorgi-Amiri et al., 2017; Shahriari et al., 2017;  Zhang et al., 2019).

Several studies propose to repurpose existing rooftop helipads, floating barges, the centers of interstate turnabouts, and top levels of parking garages, as they offer compelling space and locational advantages for urban operations. Moreover, these structures, or spaces, are rarely used in several cities (Thomaier et al., 2015; Sanyé-Mengual et al., 2016; Specht and Sanyé-Mengual, 2017). While rooftop helipads could serve as a facility for single capacity vertistops, floating barges, interstate turnabouts, and parking garages provide sufficient space for fully operational vertiports (Holden and Goel., 2016). 

\begin{figure}[!]
	\centering
	\captionsetup{justification=centering}
	\includegraphics[width=1\linewidth]{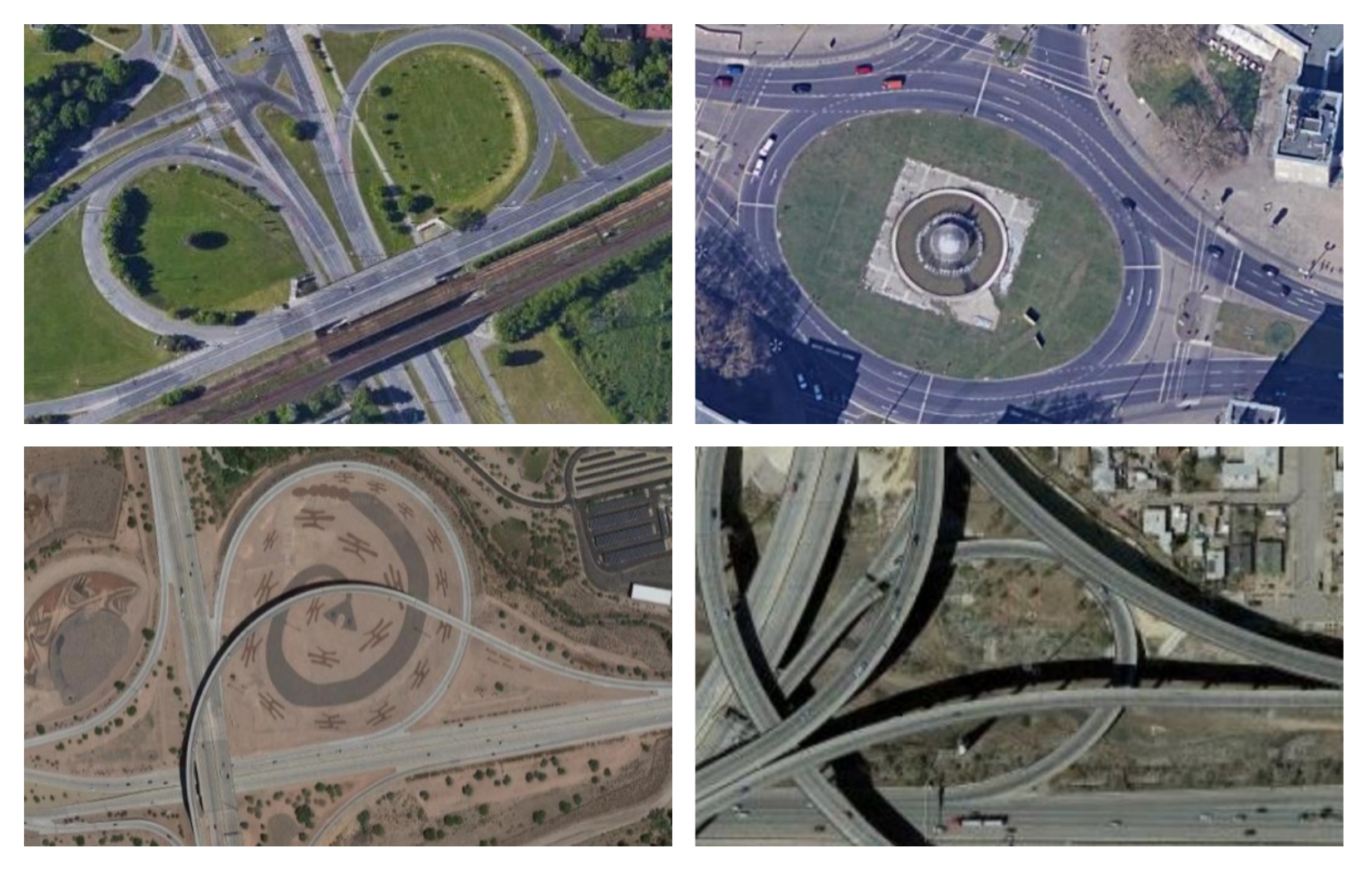}
	\caption{Examples of turnabouts that could serve as vertiports (obtained from Google Maps)}
	\label{fig:Fig3}
\end{figure}

Using such existing structures or facilities would allow faster implementation and lower overhead costs over the short-term. Interstate turnabouts, for instance, have circular turns that connect highways, as shown in Figure \ref{fig:Fig3}. Each turnaround does not typically contain structures within its large diameter (Irizarry, 2003), offering a vast amount of space that is suitable for separate air taxi vertiports. Additionally, the landscape surrounding interstate turnabouts provides a substantial allowance to support facilities for charging, repair, maintenance, and vehicle docking. Besides, it also offers an advantage to noise reduction as eVTOL noise would be blended with the that produced from existing on-road traffic (Holden and Goel, 2016). On the other hand, vertistops could be located on high rooftops, thus alleviating concerns other than proximity to other buildings during descents, however, the helipads need to be in accordance with the FAA regulations (Antcliff, 2016).

Aside from qualitative recommendations, prior research has also used analytical approaches to determine the appropriate site for a vertiport or vertistop. Rajendran and Zack (2019) developed an iterative constrained clustering approach to identify potential skyport locations for ATS in New York City (NYC). Their model detected 21 site locations in various boroughs of NYC, especially high cluster density for two locations - John F. Kennedy International Airport and South Central Park, indicating potential sites for vertiports. Similarly, Rath and Chow (2019) proposed an integer linear programming model to determine the optimal number of skyports for minimizing the travel cost associated with airport transfers. Their analysis indicated six potential sites to be adequate to serve the airport access travel needs of commuters in NYC. Although operational sites could be efficiently determined using several mathematical and heuristic approaches (Yang and Notteboom, 2016; Hammad et al., 2017; Erkan and Elsharida, 2019), these studies hardly took into consideration the safety and disturbance factors as well as infrastructure demands. Particularly for air taxi operations, it is essential to consider other parameters, such as sites that can accommodate vehicle noises, offer abundant space for safe departure and landing, and possess suitable infrastructure for operational needs (Johnson et al., 2018; Ventura Diaz et al., 2019).

\subsection{Air Taxi Configuration}
\begin{figure}[H]
	\centering
	\captionsetup{justification=centering}
	\includegraphics[width=1\linewidth]{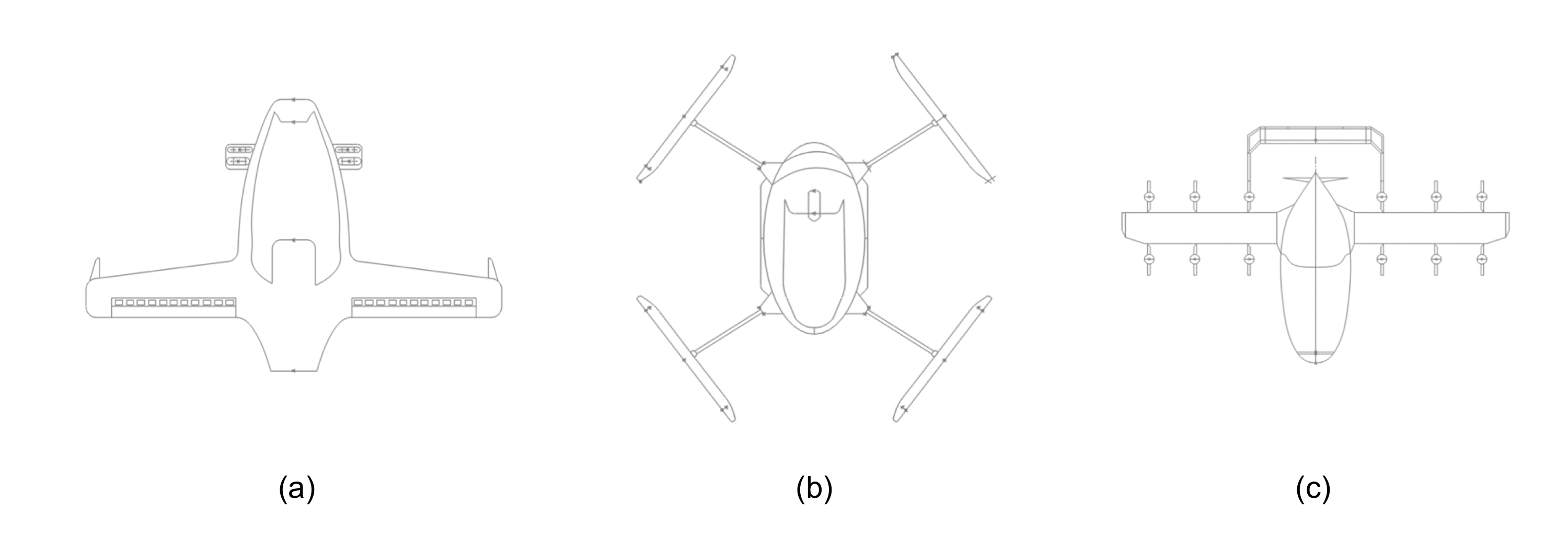}
	\caption{Illustration of three major air taxi designs - (a) vectored thrust (b) lift+cruise and (c) wingless multicopter}
	\label{fig:Fig4}
\end{figure}

Extensive research is being pursued around the globe to examine a range of air taxi configurations for designing the state-of-the-art eVTOL aircraft (Datta, 2018). Most research on the eVTOL design has resorted to the use of three primary vehicle classifications (as presented in Figure \ref{fig:Fig4}): (i) Vectored Thrust, (ii) Lift + Cruise, and (iii) Wingless Multicopter (Bacchini and Cestino, 2019). Procuring the right mix of air taxi types is a strategic choice as it has long-term impact and high capital investment. To enable informed decision-making, it is essential to comprehend these configurations as their major performance factors, such as cruise altitude, speed, flying range, and environmental impact varies across the different types (Enconniere et al., 2017; Bacchini and Cestino, 2019). 

\subsubsection{Vectored Thrust}
Vectored thrust is an aircraft’s ability to influence the direction of thrust from its engines, as shown in Figure 4(a) (Fernando, 2018). Thrust vectoring for eVTOLs is controlled through tilt-wing or tilt-rotor designs (Chauhan and Martins, 2018; Pradeep and Wei, 2018; Bacchini and Cestino, 2019). While tilt-wing designs have one or multiple propellers fixed to a wing that pivots on an axis to influence the direction of thrust, tilt-rotor designs have one or more propellers that pivot independently of the fixed-wing or structural surface to influence the direction of thrust. An advantage of vectored thrust is that it allows for higher flight speeds while in cruise configuration (Bacchini and Cestino, 2019). For instance, Airbus' $A^3$ Vahana, which is a tilt-wing eVTOL, is a full-scale prototype that operates via eight fully electric propellers mounted on forward and rear tilt wings (Polaczyk et al., 2015). Whereas, the Bell Nexus incorporates a tilt-rotor design that generates thrust via eight independent tilt-rotor propellers (Howard, 2020).

With recent advancements, compound tilt-rotor aircraft implementation has been widely considered to compensate for the gas and noise emissions from the expanding civil aviation activities. Enconniere et al. (2017) studied a performance analysis of a conceptual coaxial aircraft using simulation. While the minimum gas emissions of the conceptual vehicle occur at 80 m/s, the minimum fuel consumption occurred at 90 m/s, which is a 50\% increase from that of conventional helicopters. It is noted that fuel consumption and gas emissions decrease as altitude increases. Overall, a trade-off between gas emissions, fuel consumption, and ride duration is required (Enconniere et al., 2017). 

\subsubsection{Lift + Cruise}

An eVTOL lift + cruise design operates on independent engines for thrust and cruise capability (Figure 4(b)). Typically, fixed, vertical propellers are mounted to the top of the vehicle that generates lift similar to a helicopter, while horizontal propellers are wing-mounted to generate thrust in a forward direction for cruising. Though this design classification is not widely adopted due to their complexity, it is still a viable option for air taxi design (Deckert and Franklin, 1989; Giannini et al., 2018; Shamiyeh et al., 2018; Silva et al., 2018). Similar to vectored thrust, lift + cruise designs also allow for higher flight speeds while cruising (Shamiyeh et al., 2018; Silva et al., 2018).

Several logistics companies, such as the Cora Aero \& Kitty Hawk, developed the eVTOL with a lift + cruise design (Chauhan and Martins, 2019; Giges, 2020). This aircraft typically operates with 12 independent electric rotors mounted on its wings and one large rear-mounted rotor to generate the push. The unique safety features in this design involve electric propellers that operate independently, and in case if one fails, it will have no effect on the others, and a parachute would be released in the event that the air taxi has to land without its vertical propellers (Moore, 2020).

\subsubsection{Wingless Multicopter}
Wingless multicopter design is similar to that of a traditional helicopter operating with multiple fixed rotors (Figure 4(c)). The multicopter design classification, like the vectored thrust class, is a very popular design choice in the aerospace community behind eVTOL (Hoffmann et al., 2007; Ryll et al., 2012; Mahmoud et al., 2020). These aircraft have no wings and rely on thrust production via multiple propellers - often four or more. The advantages of a multirotor design are that the multiple rotors allow for more straightforward rotor mechanics and superior flight control, as well as reduced noise production and vibration (Lu et al., 2016). One of the most promising designs in the air taxi market is the wingless Airbus - CityAirbus. The CityAirbus is a multi-passenger, fully electric quadrotor, that utilizes autonomous flight control (Ross, 2018). The CityAirbus is designed with four ducted propellers that allow for a minimal acoustic footprint (Basset et al., 2018).

\subsubsection{Characteristics of Air Taxi Design Classifications}
Based on the description of the major air taxi configurations given by prior research (Polaczyk et al., 2015; Bacchini and Cestino, 2019; Halvorson et al., 2019; Airbus, 2020; The Vertical Flight Society, 2020), we summarize the eVTOL vehicle designs along with their characteristics and are presented in Table \ref{tab:Tab4}. We refer interested readers to Polaczyk et al. (2015) for more details. 

\begin{table}[htbp]
  \centering
  \caption{Characteristics of key air taxi design classifications}
    \scalebox{0.85}{\begin{tabular}{lccccc}
    \toprule
    \multirow{2}[4]{*}{\textbf{Characteristic}} & \multicolumn{2}{c}{\textbf{Vectored Thrust}} & \multicolumn{2}{c}{\textbf{Lift+Cruise}} & \textbf{Wingless} \\
\cmidrule{2-6}          & \textbf{Tilt rotor} & \textbf{Tilt wing} & \textbf{Fixed horizontal} & \textbf{Fixed vertical} & \textbf{Wingless} \\
    \midrule
    Example & Bell Nexus & Air bus Vahana & Kitty Hawk Cora & Cartercopter & E Hang 216 \\
    Speed (km/h) & 288   & 230   & 180   & 282   & 100 \\
    Flying Range (km) & 241   & 100   & 98.8  & 256   & 35 \\
    Seating Capacity & 4     & 2     & 2     & 4     & 2 \\
    Battery Density (kwh) & 50    & 40    & 63    & 300   & 110 \\
    Cruise Altitude (ft.) & 250   & 1,000 & 3,000 & 10,000 & 10,000 \\
    \bottomrule
    \end{tabular}}%
  \label{tab:Tab4}%
\end{table}%

It is evident that each alternate design has its own strength and weakness with regards to range, speed, passenger capacity, and environmental impact criteria. Therefore, from an operations management perspective, determining the optimal number of air taxis in each design category is essential to achieve a trade-off between service responsiveness,  operating cost, and air taxi utilization.

\section{Challenges}
This section reviews the potential unaddressed challenges associated with air taxi operations and presents the research conducted in similar avenues. 

\subsection{Ride-Matching}
\begin{figure}[!]
	\centering
	\captionsetup{justification=centering}
	\includegraphics[width=0.75\linewidth]{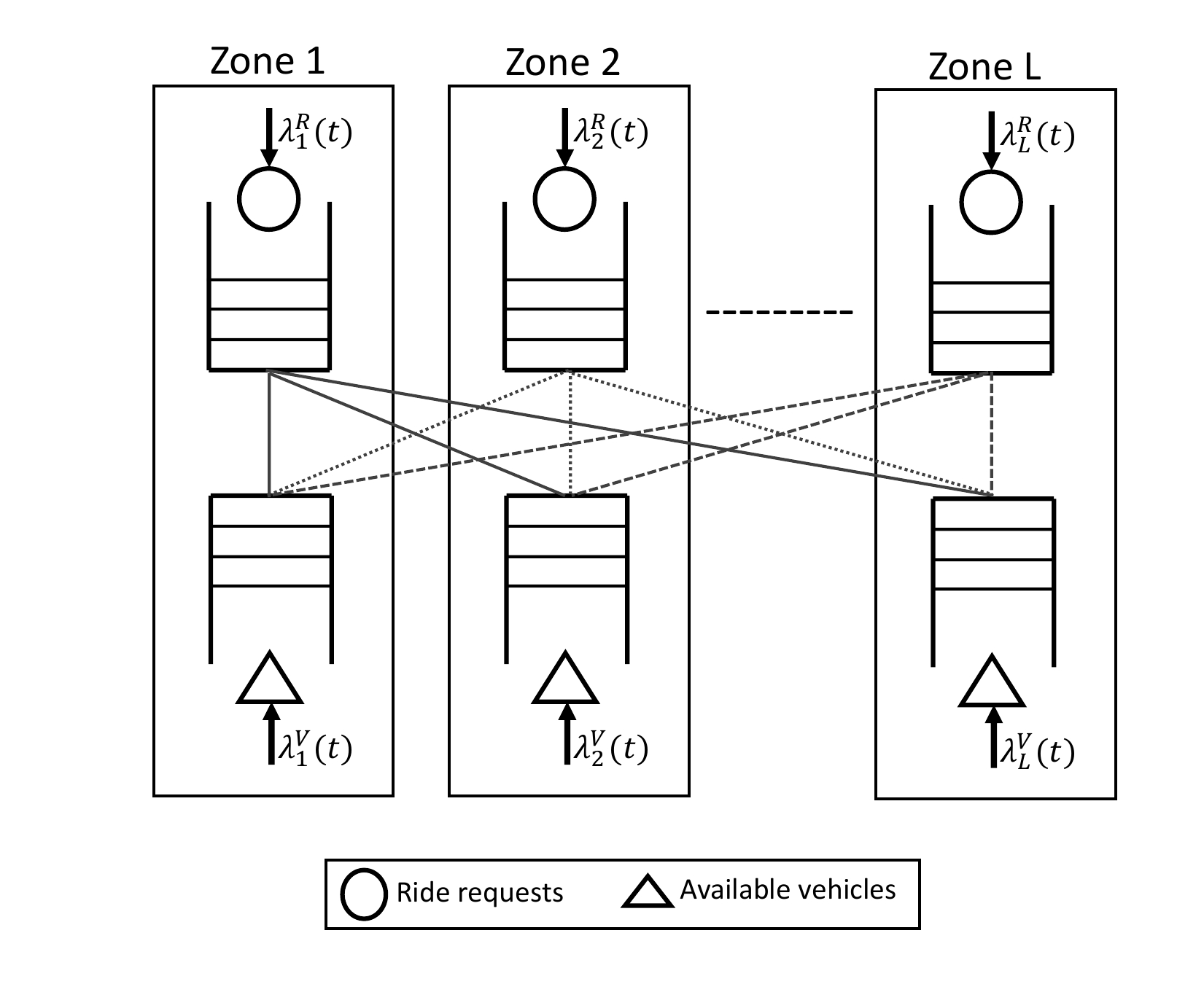}
	\caption{Schematic representation of ride-matching system}
	\label{fig:Fig5}
\end{figure}

Ride-matching is the process of assigning one of the available vehicles to a customer request. In the case of on-road ODM service, the vehicle is dispatched to the customer location for pickup. Whereas, to facilitate ATS, the passenger must travel to the assigned air taxi’s vertiport/vertistop location via on-road transport or walk. Figure \ref{fig:Fig5} illustrates a typical ride-matching system. At a given period $t$ and network location (or zone) $l$, the ride requests and available vehicles arrive at a rate of $\lambda_l^R (t)$ and $\lambda_l^V (t)$, respectively. The system controller (ride-hailing platform) matches the customer request to an available vehicle in the same zone or any other neighboring location depending on several factors, such as estimated pickup time, empty travel distance, and vehicle supply distribution across the network (Herbawi and Weber, 2012; \"Ozkan and Ward, 2020). 

For any on-demand, ride-hailing service (on-road or air taxi), efficient matching of commuters (demand) with vehicle (supply) is crucial for minimizing customer wait time and vacant/empty trips (Rajendran and Shulman, 2020). The flexibility in matching a request to a vehicle depends on the market type. In a one-sided market (OSM), the platform has only one group of customers (demand-side/commuters) as it controls the entire supply of vehicles (and drivers). On the other hand, the platform acts as an intermediary between the two groups of customers (independent drivers who own the vehicle, as well as riders) for a commission in a two-sided market (TSM). While only the customer has the capability to decline a matched trip in an OSM, both the parties (driver and rider) can cancel a paired trip in a TSM. In other words, the platform can control $\lambda_l^V (t)$ to a great extent in the OSM, whereas the vehicle supply is difficult to manage in the TSM as the independent drivers may travel to other regions searching for a ride match or leave the system. 

As the supply and demand are expected to vary over time and location, these matching systems should be automatic and dynamic to facilitate swift on-demand coordination (Agatz et al., 2012). The complexity of trip matching depends on the type of ride-hailing request, which can be broadly classified into two categories:

\begin{itemize}
	\item \textit{Single Customer Service per Trip}: The vehicle transports an individual customer from a pickup point to a destination location.
	\item \textit{Multiple Customer Service per Trip (Ridesharing)}: The vehicle accommodates numerous customer requests, which occur around the same time and have a different, but nearby, origin as well as destination locations on a trip.
\end{itemize}

Besides, another key characteristic influencing the intricacy of rider-vehicle matching is the number of trip segments (Friedrich et al., 2018). Specifically, the potential matches for a customer request rise with the number of segments, thereby increasing the computational time for enumerating all the combinations. As a result, trip matching would become a significant challenge for ATS as it is expected to have multiple segments (a combination of on-road and on-air travel) for facilitating door-to-door transportation, unlike the traditional ODM services.

The on-road segment of ATS is equivalent to the operations of existing on-demand ride services, such as Uber and Lyft. Therefore, matching algorithms of these ODM services can be adapted for transporting the rider from the pickup location to a skyport or from a station to the drop-off location. Initially, taxi scheduling studies focused on pre-booking passengers (e.g., Brake, 2008; Nelson et al., 2010). However, in recent years, dynamic approaches to request and allocate customers in real-time have been widely investigated (Agatz et al., 2012). Current matching algorithms for ODM services can be grouped into three categories – first-dispatch protocol, batching, and dynamic matching (Yan et al., 2019). The \textit{first-dispatch protocol} (FDP) assigns a customer request instantaneously to one of the idle vehicles, which has the shortest estimated arrival time at the pickup location (Feng et al., 2020). On the other hand, the \textit{batching algorithm} is a generalization of FDP, where the ride requests are accumulated for a short period and then assigned to an idle vehicle based on an optimization model (Ashlagi et al., 2018; Korolko et al., 2018). While both these approach use a myopic strategy, the \textit{dynamic matching algorithm} takes the potential future demand and supply into consideration before assigning the vehicle to a customer request (Najmi et al., 2017; Banerjee et al., 2018; Hu and Zhou, 2018; \"Ozkan and Ward, 2020). 

The efficiency of the on-road segment can be improved by involving shared rides. The inherent advantages of ridesharing are lower trip cost, conservation of fuel, reduced traffic congestion, lessened travel time, and reduced air pollution (Furuhata et al., 2013; Bai et al., 2017; Wang et al., 2017; Ma et al., 2019). Nevertheless, the matching process becomes even more complicated when incorporating shared rides. Specifically, sequencing the pickup and drop-off of multiple passengers in ridesharing complicates the decision process and makes the optimal pooling problem hard to solve (Korolko et al., 2018). Moreover, it is necessary to fuse data from multiple sources, such as GPS, and web technologies, for determining the set of customer requests that can be matched to a particular vehicle such that one or more objectives (e.g., minimize trip fare, minimize travel time) is optimized (Furuhata et al., 2013; Zhang et al., 2018; Cetin and Deakin, 2019; Sui et al., 2019). 

On the other hand, the air taxi segment (i.e., travel between vertiports/vertistops) is analogous to the Dial-a-Flight Problem (DAFP) in the transportation literature, where a set of air transport requests is matched to one of the available aviation vehicles during a specific period (Cordeau et al., 2007; Espinoza et al., 2008a). A request indicates information, such as origin and destination stations, number of passengers traveling, and earliest pickup time. Each requisition may be accepted or rejected, depending on various financial and strategic factors (van der Zwan et al., 2011). Depending on the application area, the objective of DAFP can be to minimize cost, maximize service level, or maximize profits, while satisfying the problem constraints (Cordeau et al., 2007; Espinoza et al., 2008a). A static DAFP assumes the total requests for a given period to be known in advance, while a dynamic DAFP considers the demand to occur over time. Unlike static DAFP, the decision to accept or reject the customer request for dynamic DAFP must be taken in a short period by considering the available supply and accepted requests (Cordeau et al., 2007). 

Table \ref{tab:Tab5} summarizes some of the notable works on DAFP for on-demand air transport services. Almost all studies focused on minimizing the total operational costs for the air taxi operator (e.g., Espinoza et al. 2008a; Munari and Alvarez, 2019). Besides, most prior works considered the air taxi service provider to own a homogenous fleet and have previous knowledge of the demand (static DAFP). Typically, small problem instances are formulated as integer linear program (e.g., Espinoza et al. 2008a; Munari, 2017; Campbell et al., 2020), while large cases are modeled using local search heuristics (Espinoza et al., 2008b).

While the DAFP literature shares many characteristics of the eVTOL-based ATS, there are two notable differences. First, all prior works on DAFP assume a considerable lag between trip requests and earliest departure times (LBRD), whereas ATS for urban mobility requires real-time matching and very small LBRD. Second, DAFP for traditional air transport typically considers the passenger(s) to return to the origin location upon visiting one or more sites (Reddy, 2018), but ATS is likely to have distinct pickup and drop-off locations. Therefore, these differences must be taken into consideration when adapting models from DAFP literature for the on-air segment of ATS.

\begin{table}[htbp]
	\centering
	\caption{Summary of notable works on DAFP for on-demand air transport service}
	\scalebox{0.85}{\begin{tabular}{rrrrrrr}
		\toprule
		\multicolumn{1}{c}{\textbf{Author}} & \multicolumn{1}{p{5em}}{\centering \textbf{Matching System}} & \multicolumn{1}{c}{\textbf{Objective}} & \multicolumn{1}{p{5em}}{\centering \textbf{Modeling Approach}} & \multicolumn{1}{c}{\textbf{Fleet Size}} & \multicolumn{1}{p{5em}}{\centering \textbf{Total Stations}} & \multicolumn{1}{c}{\textbf{LBRD (in h)}} \\
		\midrule
		\multicolumn{1}{l}{Espinoza et al. (2008a)} & \multicolumn{1}{c}{S} & \multicolumn{1}{c}{MOC} & \multicolumn{1}{c}{IP} & \multicolumn{1}{c}{4-8} & \multicolumn{1}{c}{17} & \multicolumn{1}{c}{$\geq$ 24} \\
		\multicolumn{1}{l}{Espinoza et al. (2008b)} & \multicolumn{1}{c}{S} & \multicolumn{1}{c}{MOC} & \multicolumn{1}{c}{OBLS} & \multicolumn{1}{c}{312} & \multicolumn{1}{c}{46} & \multicolumn{1}{c}{$\geq$ 24} \\
		\multicolumn{1}{l}{Fagerholt et al. (2009)} & \multicolumn{1}{c}{D} & \multicolumn{1}{c}{MOC} & \multicolumn{1}{c}{Heuristic} & \multicolumn{1}{c}{5 - 20} & \multicolumn{1}{c}{12} & \multicolumn{1}{c}{$\geq$ 24} \\
		\multicolumn{1}{l}{Engineer et al. (2011)} & \multicolumn{1}{c}{S} & \multicolumn{1}{c}{MFT} & \multicolumn{1}{c}{DP} & \multicolumn{1}{c}{10 - 200} & \multicolumn{1}{c}{15 - 41} & \multicolumn{1}{c}{$\geq$ 24} \\
		\multicolumn{1}{l}{Farhoudi et al. (2012)} & \multicolumn{1}{c}{D} & \multicolumn{1}{c}{MOC} & \multicolumn{1}{c}{SP} & \multicolumn{1}{c}{19 - 25} & \multicolumn{1}{c}{225} & \multicolumn{1}{c}{4 – 48} \\
		\multicolumn{1}{l}{Munari (2017)} & \multicolumn{1}{c}{S} & \multicolumn{1}{c}{MOC} & \multicolumn{1}{c}{MIP} & \multicolumn{1}{c}{25 - 30} & \multicolumn{1}{c}{35 - 52} & \multicolumn{1}{c}{$\geq$ 24} \\
		\multicolumn{1}{l}{Munari \& Alvarez (2019)} & \multicolumn{1}{c}{S} & \multicolumn{1}{c}{MOC} & \multicolumn{1}{c}{MIP} & \multicolumn{1}{c}{15 - 53} & \multicolumn{1}{c}{25 -93} & \multicolumn{1}{c}{$\geq$ 24} \\
		\multicolumn{1}{l}{Campbell et al. (2020)} & \multicolumn{1}{c}{S} & \multicolumn{1}{c}{MOC} & \multicolumn{1}{c}{IP} & \multicolumn{1}{c}{17} & \multicolumn{1}{c}{28} & \multicolumn{1}{c}{$\geq$ 24} \\
		\midrule
		\multicolumn{7}{p{48em}}{\footnotesize IP: Integer programming; MIP: Mixed integer programming; SP: Set partitioning; DP: Dynamic programming; OBLS: Optimization-based local search; S: Static; D: Dynamic; MOC: Minimize operational costs, MFT: Minimize flying time} \\
	\end{tabular}}%
	\label{tab:Tab5}%
\end{table}%

\normalsize

\subsection{Pricing Strategies for On-Demand Mobility}
Pricing refers to the method of establishing the passenger fare for an ATS. Besides base fare (a flat fee per ride), the trip cost would also be impacted by ride distance, duration, booking fees, price multiplier, promotions, and ridesharing discounts (Yang et al., 2010; Wang et al., 2016; Watanabe et al., 2016; Chen et al., 2017; Kienzler and Kowalkowski, 2017; Lam and Liu, 2017; Guda and Subramanian, 2019). Determining the pricing scheme for an ODM service is a crucial problem as it has a direct impact on supply-demand equilibrium and revenue (Saharan et al., 2020). Typically, a customer pays more than the regular price in the case of high demand and low supply (where the price multiplier is greater than 1).

To manage the imbalance between available drivers and passenger requests, prior research mostly considered \textit{real-time (or dynamic) pricing strategies} (Chen and Wilson, 2017; Qian et al., 2017; Aujla et al., 2019; Sun et al., 2019). Almost all prior works on dynamic pricing strategy for ODM focused on maximizing the profit (or revenue) of the service provider during a specific period based on the historical demand pattern. Banerjee et al. (2015) resorted to a theoretical approach by using an M/M(k)/1 queuing model to establish the threshold dynamic pricing policy, where the fare multiplier is adjusted depending on the number of available drivers. The authors found the threshold-based pricing to be robust at handling system uncertainties but yield insignificant advantage over static pricing for a single-region market. Unlike these two studies, which factored only the price-sensitive nature of customers, Zha et al. (2017) considered both trip fare and customer waiting time to affect the demand. They concluded that dynamic pricing could benefit both the driver and platform in a two-sided market as opposed to static pricing, whereas the customers only benefit during the off-peak hours. 

While the aforementioned works considered a temporal supply-demand imbalance, other works focused on spatial pricing strategy, in which the ride fare varies across the ODM network locations depending on the customer demand (Zha et al., 2018; Bimpikis et al., 2019). Zha et al. (2018) used a matching algorithm to investigate spatial pricing and observed it to avoid inefficient supply state while the customers may be overcharged. Therefore, the authors proposed a regulation on the commission rate to facilitate spatial pricing and customer welfare. Likewise, Bimpikis et al. (2019) considered an ODM platform serving a set of geographical locations and explored the impact of spatial price discrimination. They concluded that a balanced demand pattern across the network locations would maximize the profit. Besides, they also found origin-based pricing to be optimal for an unbalanced network demand.

Furthermore, few studies on dynamic pricing took into account both temporal and spatial mismatch between supply and demand. Luo and Saigal (2017) used a continuous-time continuous-space approach to reduce the computational burden of handling the spatiotemporal pricing problem and developed a dynamic programming model to maximize revenue. Ma et al. (2018) conducted an empirical study on ODM pricing and found the spatiotemporal pricing scheme to achieve significantly higher social welfare as well as driver time-efficiency than a myopic pricing policy. Instead of establishing or evaluating a dynamic pricing strategy like most previous research, Battifarano and Qian (2019) developed a machine learning method (log-linear model with L1 regularization) and leveraged relevant historical data to the predict price multiplier ahead of time. 

On the other hand, the real-time pricing literature on a one-sided ODM market is limited (Lei et al., 2020). Qiu et al. (2018) focused on establishing a dynamic pricing strategy by considering the distribution of upcoming trip requests and commuter preferences to be known in advance. Instead of assuming the demand distribution to know apriori, Lei et al. (2020) optimized the path-based pricing and vehicle dispatching decisions by considering a spatiotemporally unbalanced demand. Their proposed pricing strategy resulted in significantly higher revenue than myopic policies in various stochastic settings. Due to the prevalence of real-time pricing in the ODM market, customers are already accustomed to this strategy (Banerjee al., 2015; Kienzler and Kowalkowski, 2017; Korolko et al., 2018), and might be willing to accept it for ATS.

While dynamic pricing is the most popular strategy, a similar technique known as \textit{time-of-use (TOU) pricing} is also feasible, where the rate structure is altered depending on when a transport service is used. Unlike dynamic adjustments, the price multiplier under TOU is predetermined for different time periods. Even though TOU pricing is ubiquitous in the utility industry (Nelson and Orton, 2013; Wang et al., 2015; Lin et al., 2019; Kumari et al., 2020), it is not extensively studied for the ODM market. Both dynamic and TOU pricing schemes burden the commuter with fare increases during high demand periods. To overcome this drawback, a \textit{subscription-based pricing strategy} can be adopted as it protects customers from high price fluctuations by charging a fixed fee per time period in advance (Kung and Zhong, 2017; Matyas, 2019). Based on the subscription plan, a user may be eligible for a discounted fare or a fixed number of rides during the subscription period. To evaluate its effectiveness, popular ODM platforms, such as Uber and Lyft, recently unveiled subscription services in selected US locations (Bilen, 2018; Baker, 2018; Wheeler, 2019). However, due to the novelty of this pricing scheme in ODM service, it is rarely studied in the literature.

\begin{figure}[H]
	\centering
	\captionsetup{justification=centering}
	\includegraphics[width=1\linewidth]{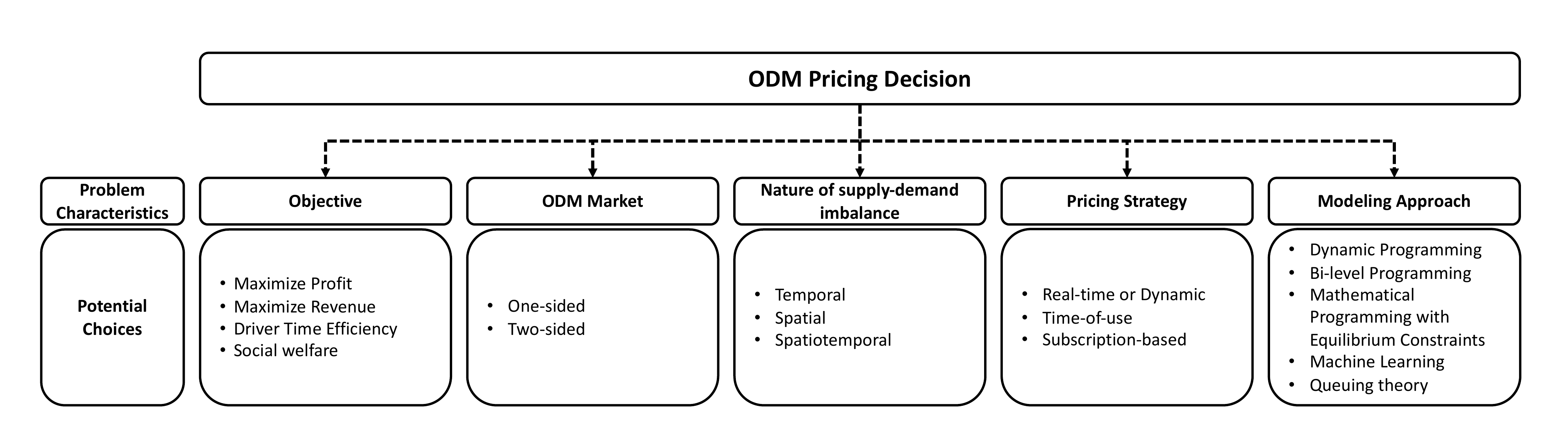}
	\caption{Taxonomy of important characteristics and potential choices for ODM pricing decisions}
	\label{fig:Fig6}
\end{figure}

Figure \ref{fig:Fig6} provides a summary of key characteristics associated with ODM pricing decisions along with the potential choices based on the review of relevant literature. Air taxis are likely to operate as a one-sided market and could experience spatiotemporal supply-demand imbalance. The rate structure for ATS can be determined by capitalizing and adapting the extensive research on dynamic pricing.

\subsection{Fleet Maintenance}
To ensure the operational safety and reliability of aviation vehicles, specific maintenance programs are mandated by the Federal Aviation Administration (FAA) in the US. Typically, the maintenance process in the aviation sector involves a series of inspections along with unscheduled repairs (El Moudani and Mora-Camino, 2000; Sriram and Haghani, 2003). These inspections may take place frequently and depends on flight utilization and the number of takeoff and landings since the previous check (Bugaj et al., 2019; Sanchez et al., 2020). 

In traditional aviation services, four different types of maintenance checks (A, B, C, and D) are performed to comply with FAA policies (Sriram and Haghani, 2003). Type A inspections must occur about once a week, focusing on the assessment of all critical components, such as landing gear, engines, and control surfaces. The next major check is Type B, which involves the visual inspection and lubrication of aircraft parts. On the other hand, categories C and D checks are relatively infrequent and are conducted once every 1-4 years. Whereas, air taxi electric motors are expected to have a time between overhaul (TBO) of 10,000 hours, with daily visual inspections and preventive maintenance being performed at an interval of 100 hours (Holden and Goel, 2016). These downtimes could significantly affect the vehicle's ability for serving customers. Therefore, while developing the air taxi maintenance schedule, it is essential to consider factors, such as demand, vehicle utilization, TBO, visual inspection, and minor maintenance checks. Moreover, the air taxi maintenance and labor cost are anticipated to constitute about 22\% of the VTOL direct operating costs (Holden and Goel, 2016). Due to these significant attributes, fleet maintenance, repair and overhaul (MRO) is considered as one of the key decisions for efficient aviation services (Papakostas et al., 2010; Atli and Kahraman, 2012; Liu et al., 2019).  In addition, these periodic maintenance would have to be conducted on a large fleet of vehicles, especially when the air taxi market matures. The vertiport crews would be responsible for charging vehicles, performing pre-flight checks, avionics system checks before departure, and maintaining overall vehicle cleanliness (Holden and Goel, 2016). 

As a result of the unique characteristics of the aviation industries, such as high operating costs,  strict regulations, complex planning, and coordination processes, decisions related to fleet maintenance has been widely studied in the literature (Cordeau et al., 2001; Mercier et al., 2005; Díaz-Ramírez et al., 2014; Cui et al., 2019; Deng et al., 2020). Two of the major issues in the aircraft maintenance scheduling problem are the ever-changing pattern of flight schedules and different aircraft service requirements (Chan et al., 2005). Prior research has tackled these problems using both optimization models and heuristic approaches (Cordeau et al., 2001; Chan et al., 2005; Ezzinbi et al., 2014; Cui et al., 2019; Sanchez et al., 2020). 

Sriram and Haghani (2003) intended to minimize the MRO cost by utilizing an integer-programming (IP) model for maintenance scheduling. Due to the computational complexity of the IP model, a heuristic algorithm was proposed to obtain a solution quickly and efficiently. Similarly, Chan et al. (2005) proposed the use and application of an intelligent engine to develop a set of computational schedules for the maintenance of vehicles. The authors utilized genetic algorithms and concluded that their developed approach provided efficient and effective maintenance schedules. Qin et al. (2017) further extended the research on aircraft maintenance scheduling considering the parking stand planning problem, in which decisions also have to be made to determine the parking layout for efficient maintenance operations. These decisions also have to be made in ATS while designing the layout of vertiports, a facility comprising of customer pickup/drop-off, maintenance and charging stations, as well as, docking sites. A list of key findings from prior works on aircraft maintenance, that could also be applicable for air taxi MRO operations is summarized in Table \ref{tab:Tab6}.

\begin{table}[H]
	\centering
	\caption{Summary of selected literature on aircraft maintenance}
	\scalebox{0.7}{\begin{tabular}{lp{10em}p{15em}p{20em}}
		\toprule
		\textbf{Article} & \multicolumn{1}{c}{\textbf{Approach}} & \multicolumn{1}{c}{\textbf{Objective}} & \multicolumn{1}{c}{\textbf{Key Findings}} \\
		\midrule
		Sriram and Haghani (2003) & IP and heuristic & Minimizing maintenance cost & Heuristic methods are essential for solving the maintenance scheduling problem given a schedule. The quality of the heuristic approach solution significantly depends upon the number of different combinations of aircraft and node orders under consideration. \\
		Chan et al. (2005) & Metaheuristic & Maximize the utilization of ground support vehicles & The application of an intelligent system is efficient and effective to generate schedules for vehicle maintenance considering the variation of flight schedules, developing contingency plans to handle any unexpected situations. \\
		Qin et al. (2017) & MILP & Maximize profit & The proposed approach will assist independent aircraft maintenance companies in increasing the efficiency of the hangar space, while simultaneously improving customer service level. \\
		Cui et al. (2019) & IP and heuristic & Minimize the number of aircraft and total remaining flying time & Models considering the flight delay probability can result in a robust flight scheduling plan, considering the addition of cushion time \\
		Liu et al. (2019) & Simulation & Reduce maintenance costs, improve service level, and reliability & The operational support can be improved by integrating aircraft’s condition, strategy, planning, and cost, into the model \\
		Deng et al. (2020) & DP & Minimize the wasted interval between checks & By considering the objective of minimizing the wasted interval between checks, other measures, such as the maintenance cost is reduced, while simultaneously increasing the aircraft availability. \\
		Sanchez et al. (2020) & Multi-objective MILP and iterative algorithm & Minimize the number of maintenance regulation violations & Efficient resource allocation is guaranteed when there exists cooperation among competitors for workshop resource sharing. \\
		\bottomrule
		\multicolumn{4}{p{50em}}{\footnotesize DP: Dynamic programming; IP: Integer programming; MILP: Mixed integer linear programming} \\
		
	\end{tabular}}%
	\label{tab:Tab6}%
\end{table}%

\subsection{Pilot Training and Scheduling}
Subsequent to the fuel cost, expenses associated with the manpower constitute the second largest cost component in the air transportation industry (Ball et al., 2007). Although ATS is not projected to function with flight attendants, the vehicles are required to be operated by pilots; nevertheless, in the long run, they are expected to be fully autonomous. Pilot training and scheduling is a challenging workforce planning problem faced by the commercial aviation sector (Qi et al., 2004). Besides, to become an air taxi pilot is extremely time-intensive and requires over 2000 hours of training (Holden and Goel, 2016). Moreover, the demand for air taxi pilots is predicted to be high, specifically during the initial stages of ATS implementation (Holden et al., 2018). 

Pilot training and certifications fall under FAA Section 14 Code of Federal Regulations (CFR) Part 135 (Federal Aviation Administration, 2019), which covers commercial ATS. Part 135 presents the following requirements for pilots: (i) 500 hours of pilot in command (PIC) experience, (ii) 500 hours for visual flight rules (VFR), (iii) 1200 hours for instrument flight rules (IFR), and (iv) 50 hours of PIC time in powered lift (helicopter or VTOL) vehicle. Hence, obtaining certification for a part 135 classification license could be time-consuming. To combat this, air taxi pilot augmentation technology is being proposed to reduce the training requirements of operators, thus also curtailing the overall training hours (Holden and Goel, 2016). For instance, the current air taxi autonomous system design has the potential ability to control engine failures, navigational equipment failures, and even diminished vehicle conditions (Board, 2005; Staplin et al., 2018).

While pilot training is of crucial importance for guaranteeing system safety, effective scheduling decisions for achieving a trade-off between workforce over-burden and resource idle time also have to be made. Prior works have concluded that scheduling practices followed by the aviation industry play an important role in addressing fatigue issues in pilots (e.g., Gregory et al., 2010). Once pilots are trained, their flight time may not exceed eight hours during 24 consecutive hours, as per the FAA regulations. While pilot scheduling models have been developed in the literature over many decades (e.g., Stojković and Soumis, 2001; Chen et al., 2016), these algorithms may not be directly applicable for ATS, as the air taxi manpower scheduling has several unique attributes associated with it. For instance, air taxis are short distance aviation services compared to long-haul flights that operate for at least a couple of hours. Also, as a significant number of air taxis are expected to take off and land at metropolitan cities simultaneously, factors, such as pilot briefing and debriefing times, flight minimum connection time varies for ATS. Moreover, it is necessary to build a pilot scheduling algorithm that optimally matches the pilot to air taxis considering restrictions, such as the maximum duty duration. 

\begin{figure}[h]
	\centering
	\scalebox{0.9}{\begin{tikzpicture}[align=center,
	>=latex,
	font=\sffamily
	]
	\node[font=\large\sffamily, minimum height=4cm,minimum width=15cm, yshift=-0.5cm] (P1) {\textbf{Phase 1:} Determining the number of pilots required per shift\\ \begin{flushleft} Objective: Min $\displaystyle Z_1 = \sum_i x_i$ \\ {s.t.}\\ $\displaystyle x_1 + x_2+...+x_i+x_{\mathcal{N}-e+i+1}+...+x_{\mathcal{N}} \geq P_i \quad \forall i \in \mathcal{N} \, \text{and} \, i \leq |e| $\\ 
		$\displaystyle x_{i-e}+x_{i-e+1}+...+x_i \geq P_i \quad \forall i \in \mathcal{N} \; \text{and} \; i > |e| $	
		
		\end{flushleft}
	} ;
	\node[font=\large\sffamily, below of= P1,minimum height=1cm, yshift=-3.8cm] (P2) {\textbf{Phase 2:} Pilot Scheduling \\ \begin{flushleft} 
		Objective: Min $Z_2 = \sum_j \sum_h W_h \times Y_{P_{h,j},j}$\\ {s.t.}\\ $\displaystyle \sum_j Y_{ij} = x_i \quad \forall i \in \mathcal{N}$ \\
		$\displaystyle \sum_i Y_{ij} = 1\quad \forall j \in \mathcal{J}$

		\end{flushleft}
	}; 
	
	\draw [arrow] (P1) -- (P2);

	\end{tikzpicture}}
	\caption{Overview of the two-phase approach for pilot scheduling}\label{fig:2P}
\end{figure}
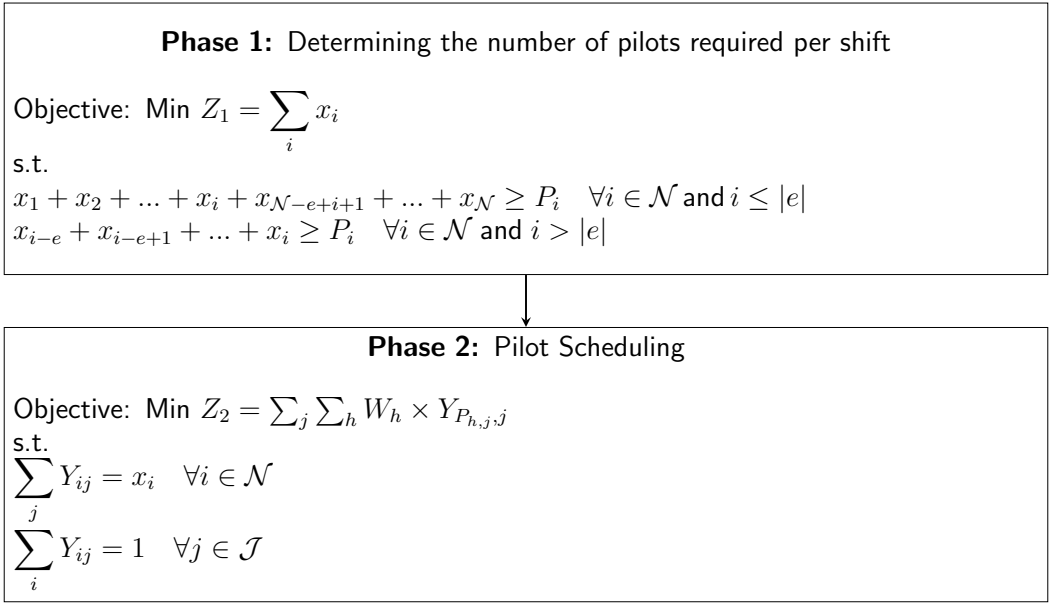

A basic framework to address the problem of pilot selection and scheduling using a two-phase approach is presented in Figure \ref{fig:2P}. Typically, the pilot schedule is divided into $\mathcal{N}$ slots, where $P_i$ pilots are required in each slot $i \in \mathcal{N}$. Besides, operators who begin their shifts in time slot $i \in \mathcal{N}$ must end their shifts after $e$ slots. Given these conditions, if $x_i$ is the number of pilots who begin their shifts in time slot $i$, then the objective is to minimize the total number of pilots, such that the requirement at each slot is fulfilled. Phase-2 involves the assignment of $j \in \mathcal{J}$ pilots to $i \in \mathcal{N}$ slots, given $x_i$ as an input. Each pilot $j$ specifies the $h^\text{th}$ preference start slot (e.g., if pilot \#9 has the first priority to start his/her shift at slot \#2 and second priority as slot \#3 – then $P_{1,9} = 2$ and $P_{2,9}=3$). Based on the input from the decision-maker on the value of $W_h$, the weight associated with preference $h$ (note: $W_1>>W_2>>W_3…$), and $x_i$ from phase 1, the scheduling algorithm obtains $Y_{ij}$, a binary variable that takes the value 1, if pilot $j$ is assigned to slot $i$ as his/her start shift, with the objective of maximizing pilot preference.

The framework provided in Figure \ref{fig:2P} is elementary, and with the implementation of ATS, several decisions would have to be made in an integrated manner instead of adapting a sequential approach. For instance, prior studies have simultaneously considered aircraft maintenance and workforce scheduling problem and have proven to obtain better performance measures (e.g., reduced cost and time) than the serial approach (Cordeau et al., 2001; Mercier and Soumis, 2005; Díaz-Ramírez et al., 2014). 
\\

\section{Opportunities for Future Research Directions}
Based on the review of the current developments and challenges in the literature pertaining to air taxi systems, we observe the following research areas to be under-explored.

\begin{enumerate}[label=\roman*.]
\item \textbf{Optimizing Fleet Procurement}: An excess inventory of air taxis would result in high procurement and maintenance costs. On the other hand, the unavailability of these vehicles could lead to long customer wait times or rejecting ride requests, thus causing unsatisfied passengers. Therefore, the availability of air taxis at the right time at the right place is essential to maintain a specified customer service level, while minimizing cost. Besides, as discussed in Section 3.2, several air taxi designs are proposed in prior research with varying flying range, speed, passenger capacity, and environmental impact. Hence, the air taxi domain is further associated with the challenge of procuring the optimum fleet of vehicles in each configuration to reduce cost. 

Another crucial challenge associated with air taxi vehicles is their long production time. Given the significant manufacturing time of air taxis, advance demand estimation before the introduction of this aviation service is extremely crucial, unlike the traditional on-road ODM services. Therefore, as a potential future work, studies could investigate the development of integrated models for demand estimation and air taxi fleet procurement. 

\item \textbf{Determining Station Capacities}: Regarding the air taxi facility selection problem, prior works have attempted to determine the location of vertiports and vertistops based on the customer demand volume in metropolitan cities (e.g., Rajendran and Zack, 2019; Rath and Chow, 2019). However, it is essential to determine the size of these facilities, particularly vertiports. As discussed earlier, vertiports are larger sites, comprising of maintenance, charging, docking, and multiple takeoff and landing facilities. Hence, before making a site location decision, it is essential to determine in advance the number of sub-stations required in each of these categories, as well as the layout of these facilities. Thus, future research could involve developing models for the vertiport selection problem considering these capacity constraints. 

\item \textbf{Multi-Criteria Decision Analysis (MCDA) for Vertiport/Vertistop Selection}: Besides the customer demand, there are several conflicting criteria that are involved in selecting a facility for efficient network operations (Farahani et al., 2010). A few of the air taxi-related critical criteria could include the operating cost, ease of public access, population coverage, and feasibility of establishing a site. For instance, a site that has the potential to serve a major customer base might be located in a densely-populated region, which could result in high operating and rental costs. Similarly, the feasibility of setting up a facility might be low in a highly-congested region, such as a tourist spot, which, however, has the potential to serve a large population of customers. 

In recent years, several MCDA tools have been widely explored. These methods enable decision-makers to obtain a trade-off between different objectives, considering the diverse alternatives. The order of importance of each criterion is captured either by assigning weights to the objectives or by directly obtaining the prioritization from the DM. Hence, future research could focus on developing MCDA models to rank the list of potential vertiport/vertistop locations for ATS. 

\item \textbf{Pricing Strategy for Air Taxi Operations}: Determining the optimal pricing strategy for regular taxis is extensively discussed in the literature. Even though ATS is similar to ground taxi operations with regards to certain characteristics (e.g., on-demand services, real-time routing), the pricing strategies developed for the latter are predominantly for a two-sided market. Since air taxi service providers are likely to control the fleet of vehicles, future research could adapt existing pricing strategies for a one-sided  market.


Moreover, penetration pricing strategies that are primarily used for new services during the initial phase of implementation can also be considered to increase customer demand. During the initial stages, in addition to considering factors, such as the day of the week, time of the day, and location, research pertaining to the ATS pricing strategy could also consider the influence of loyalty rewards program and discounted fare for longer rides. Moreover, since ATS is expected to launch in urban areas, specific pricing models could also be developed for special events, such as football matches.

\item \textbf{Air Taxi Maintenance Scheduling}: Unlike regular ridesharing taxi services in which each car is maintained by independent entities, air taxis have to be serviced by the logistics company. Research on demand estimation is not only a key component for fleet procurement decisions, but is also essential for developing aircraft maintenance schedules. Moreover, it is necessary to achieve a balanced utilization of resources, such as manpower and equipment.  

\item \textbf{Dynamic Routing of Air Taxis}: In recent years, with the advancement of technology and having the ability to trace the Global Positioning Systems (GPS) location, several dynamic routing algorithms are being proposed (Wang et al., 2020). These models attempt to develop cost-effective real-time solutions, based on the number of operating air taxis, location of the vehicles, aircraft charging and maintenance requirements, and customer details (e.g., pickup/dropoff location, time of pickup, willingness to rideshare, number of passengers). With the launch of ATS, future research could focus on developing efficient dynamic routing algorithms in the cyber-physical system setting a balance between customer waiting time, vehicle idle time, and travel cost. 

After every passenger dropoff, dynamic decisions have to be made on the state of the air taxi – whether it must be idle, undergo maintenance check or visual inspection, obtain battery charge, or pick up customers from other sites. If the air taxi is operational, then another decision has to be made on the location of the vertiport/vertistop in which the vehicle must pickup customers. In case if the vehicle needs to undergo an inspection, then the system has to decide the location of the vertiport where the maintenance has to be conducted, based on factors, such as resource (e.g., maintenance workforce, equipment) and space availability. 

\item \textbf{Integration of Ground and Air Transportation Scheduling}: To ensure door-to-door services for customers, ATS is complemented by ground transportation for the first- and last-mile delivery. As discussed in Section 1.2, customers are picked up by car from their actual pickup location to the source vertiport/vertistop (or walk if the station is at close distance). The drop-off event also follows a similar sequence of operations. To offer cost-effective network operations, it is essential to consider a joint scheduling approach to coordinate both modes of transportation. The integration of the matching and scheduling problem for all the segments in ATS can be viewed as a variation of the dynamic Dial-a-Ride Problem with Transfers (DARPT), where a set of dynamic customer requests must be transported from their origin to destination via one or more intermediate transfers (Masson et al., 2014). The passenger dropped at a transfer point will be transported to the next stop or destination by another vehicle. Depending on vehicle capacity, customers traveling to similar destinations may share the ride. The passengers are matched to one or more vehicles, which are then sequenced to optimize an objective, such as operational cost or total ride time (Masson et al., 2014). 

Similar to ATS, a passenger in DARPT is likely to be transported via multiple segments by multiple vehicles along with other riders. Nevertheless, there are some critical distinctions between DARPT and ATS operations. The DARPT considers a fixed supply of vehicles, but the on-road segment of ATS has a dynamic vehicle availability due to independent drivers. Moreover, DARPT assumes the vehicles to depart from one or more central locations, whereas the vehicles can be located anywhere in the network for the ATS. Finally, DAPRT is modeled for low demand transport applications (e.g., para-transit, travel between home and health centers), while ATS is expected to operate on a large-scale for urban transport. Therefore, it would be interesting to extend the literature pertaining to DAPRT for ATS. Besides integration, future works could also explore the possibility of incorporating multiple eVTOL segments/transfers (similar to the airline hub-and-spoke network) to achieve overall cost effectiveness.


\item \textbf{Leveraging Mobility as a Service (MaaS) Facility for ATS}: The incorporation of other intermodal transportation with ATS could also be examined in the future. For instance, research related to Mobility as a Service (MaaS) attempts to deliver affordable ODM commutes leveraging existing public transportation infrastructure (Stopka et al., 2018). Therefore, it will be interesting to develop joint scheduling algorithms to increase passenger travel options (e.g., bus $\rightarrow$ air taxi $\rightarrow$ subway, tram $\rightarrow$ air taxi $\rightarrow$ car) at a reduced cost.

\item \textbf{Short-term Demand Prediction}: In the case of new or impending ODM services like air taxis, a qualitative approach for demand estimation is warranted. However, once operational, air taxis provide a deluge of trip data, such as GPS traces, time of pickup/drop-off at every vertiport location, weather conditions, total passengers traveled, and battery efficiency, as air taxis are IoT-enabled and fitted with sensors (Rajendran and Shulman, 2020). This information can be capitalized to develop an effective dispatching system and artificial intelligence-enabled demand prediction models (Suh and Ryerson, 2019). It would be particularly interesting to use data analytic techniques, such as machine learning, to predict the demand for air taxis using several ride-related factors (e.g., day of the week, time of the day, location), and environment-related factors (e.g., temperature, weather condition). 

\item \textbf{Harnessing air taxis for supply chain logistics}: In recent years, last-mile deliveries using unmanned aerial vehicle (UAV) or drone has gained popularity, owing to its cost- and time-effectiveness (Kirschstein, 2020; Salama and Srinivas, 2020). As a potential scope of improvement, studies can investigate a hybrid system involving the operation of multiple UAVs in tandem with air taxis. In particular, the last-mile delivery of low-weight goods could be parallelized by dispatching drones from a single air taxi simultaneously to the nearby sites. The integration of drones with air taxis could be particularly advantageous after the onset of a disaster, during which traditional transportation infrastructure (such as roads and bridges) could be damaged, and hence, a truck-only delivery or a truck-drone system would be challenging.


\item \textbf{Pilot Training and Scheduling}: As discussed in Section 3.4, it is crucial to determine the optimal number of pilots required for efficient ATS, given their long training time and high manpower cost. Moreover, during the near-term implementation of ATS, there might be several new FAA guidelines becoming effective. Therefore, a more in-depth exploration of the development of pilot assignment models, incorporating the FAA requirements, is necessary. Also, similar to ride-matching algorithms that allocate customers to air taxis, models that assign pilots to air taxis considering attributes, such as remaining shift time, aircraft maintenance schedule, vehicle charge availability, could also be proposed. MCDA tools could be leveraged to obtain a pilot schedule that establishes a trade-off between resource overburden and idle time.

\item \textbf{Pilotless Air Taxi Systems:} As discussed earlier, expenses associated with manpower is the second-largest cost incurred by the aviation sector. Moreover, the presence of a pilot also reduces the load capacity, which, in turn, decreases the vehicle efficiency. As a result, pilotless VTOL flights, which are operated using an onboard computer interfacing with a ground control center, are being proposed in recent studies (Ma, 2017). Therefore, research on integrating the aforementioned future directions for pilotless VTOL designs could also be explored by taking into account factors, such as customer's willingness-to-fly in an unmanned system.

\end{enumerate}

\begin{figure}[H]
	\centering
	\captionsetup{justification=centering}
	\includegraphics[width=0.75\linewidth]{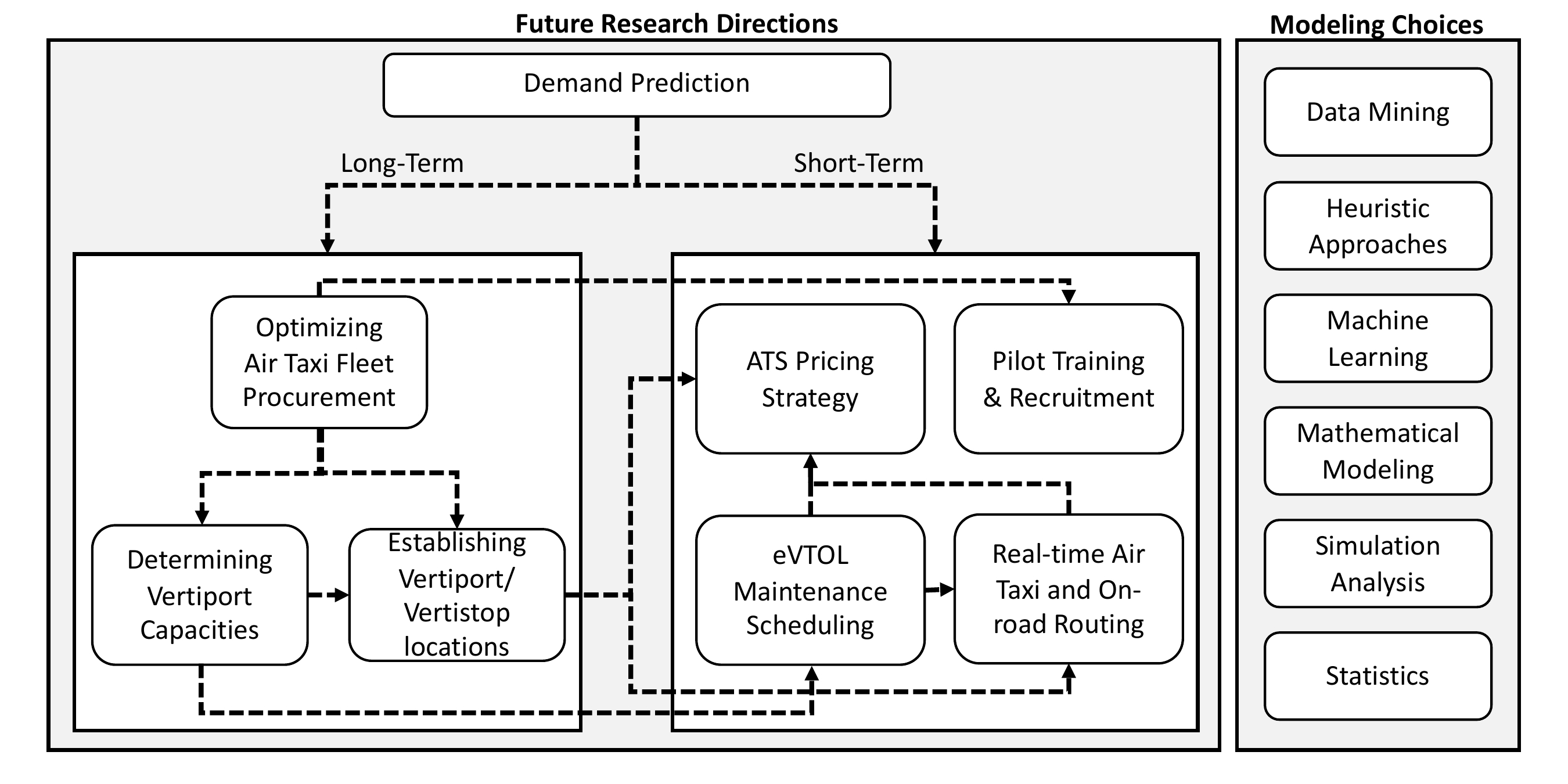}
	\caption{Inter-relationship between different air taxi network decisions}
	\label{fig:Fig8}
\end{figure}

Figure \ref{fig:Fig8} shows the inter-relationship between the different air taxi decisions discussed above from an operations management perspective. The  estimation of the long-term demand during the introductory phase of ATS would be a key input for determining the fleet required to maintain a specified customer service level. Based on the number of air taxis operating and the estimated customer volume, the location of the vertiport facility, its capacity and layout, have to be decided. On the other hand, the short-term demand estimation could have a direct positive influence on the pricing strategy. Similarly, the air taxi maintenance schedule would depend on the demand pattern across different times of the day and days of the week. Moreover, the real-time routing decisions would be impacted by the number of air taxis operating and undergoing maintenance, and station locations.  

Several modeling approaches can be leveraged to tackle these future research opportunities in the air taxi domain. Typically mathematical models, such as linear programming, mixed-integer programming, and dynamic programming, are used for optimizing decisions like pricing, fleet procurement, facility location, maintenance scheduling, and workforce allocation (Hammad et al., 2017; Corberán et al., 2020; Yun et al., 2020). Likewise, to obtain scalable and computationally efficient solutions for large-scale problems, prior research has also used simulation (e.g., discrete-event systems simulation and agent-based simulation), heuristic approaches (e.g., neighborhood search algorithms, greedy randomized adaptive search procedure) and meta-heuristic algorithms (e.g., genetic algorithms, simulated annealing). In recent years, data mining and machine learning algorithms (neural networks, support vector machines) are evolving due to the rapid emergence of big data (Choi et al., 2018).  

The impact of data science, particularly in air transportation and logistics network operations, is monumental (Choi et al., 2019; Li et al., 2019; Yu et al., 2019).  Especially with the current ability to obtain real-time data from a diverse set of sources on flight performance, air traffic conditions, and weather (Chung et al., 2020), dynamic air taxi operational decisions could be made. Notably, big data analytics using artificial intelligence and machine learning, can be used for predominantly making short-term decisions, such as preventive air taxi maintenance, real-time trip pricing, and dynamic routing and dispatching decisions.

\section{Concluding Remarks}
In an effort to ameliorate the current traffic congestion in urban areas, air taxi service (ATS), an aerial on-demand transport, is anticipated to be launched in the forthcoming years. In this study, we reviewed numerous survey papers in the air taxi and associated fields. We provided the current state-of-the-art research on air taxi systems and surveyed recent studies in demand prediction, air taxi network design, and vehicle configurations. We next identified and reviewed the potential unaddressed challenges (ride-matching, pricing strategies, vehicle maintenance scheduling, and pilot training and recruitment) associated with air taxi operations. After examining prior works and discussing the associated challenges, we have presented the future research agenda that comprises of topics under each of the core areas that we have investigated. We strongly believe that the discussions provided in this paper will be useful to both academicians and practitioners to understand the air taxi operations, the research challenges that need to be considered from an operations management perspective, and the potential works that could be explored in the forthcoming years.

\section*{References}
\begin{hangparas}{.15in}{1}
\small
\setstretch{1.0}
\setlength{\parskip}{0.45em}

References
Aarhaug, J., \& Skollerud, K. H. (2014). Taxi: different solutions in different segments.

Aaseng, G. B. (2001, October). Blueprint for an integrated vehicle health management system. In 20th DASC. 20th Digital Avionics Systems Conference (Cat. No. 01CH37219) (Vol. 1, pp. 3C1-1). IEEE.

Airbus (2020). Airbus looks to the “underutilised sky” for urban mobility. (2020). Retrieved 29 May 2020, from https://www.airbus.com/newsroom/stories/airbus-looks-to-the-underutilised-sky-for-urban-mobility.html

Alonso-Mora, J., Samaranayake, S., Wallar, A., Frazzoli, E., \& Rus, D. (2017). On-demand high-capacity ridesharing via dynamic trip-vehicle assignment. Proceedings of the National Academy of Sciences, 114(3), 462-467.

Amiri-Aref, M., Farahani, R. Z., Hewitt, M., \& Klibi, W. (2019). Equitable location of facilities in a region with probabilistic barriers to travel. Transportation Research Part E: Logistics and Transportation Review, 127, 66-85.

Antcliff, K. R., Goodrich, K., \& Moore, M. (2016, March). NASA silicon valley urban VTOL air-taxi study. In On-demand mobility/emerging tech workshop, Arlington (Vol. 7).

Ashlagi, I., Burq, M., Jaillet, P., \& Saberi, A. (2018). Maximizing efficiency in dynamic matching markets. arXiv preprint arXiv:

Atli, O., \& Kahraman, C. (2012). Aircraft maintenance planning using fuzzy critical path analysis. International Journal of Computational Intelligence Systems, 5(3), 553-567.

Aujla, G. S., Kumar, N., Singh, M., \& Zomaya, A. Y. (2019). Energy trading with dynamic pricing for electric vehicles in a smart city environment. Journal of Parallel and Distributed Computing, 127, 169-183.

Bacchini, A., \& Cestino, E. (2019). Electric VTOL configurations comparison. Aerospace, 6(3).


Bai, C., Wang, X. L., Liu, W., \& Yang, H. (2017). A Long Term Impact of Ridesharing on Private Car Ownership.

Baker, M. B. (2018). Lyft Launches Monthly Subscription Service. Retrieved June 3, 2020, from \href{https://www.businesstravelnews.com/Transportation/Ground/Lyft-Launches-Monthly-Subscription-Service}{\nolinkurl{https://www.businesstravelnews.com/Transportation/Ground/Lyft-Launches-Monthly-Subscription-Service}}

Ball, M., Barnhart, C., Nemhauser, G., \& Odoni, A. (2007). Air transportation: Irregular operations and control. Handbooks in operations research and management science, 14, 1-67.

Banerjee, S., Kanoria, Y., \& Qian, P. (2018). State dependent control of closed queueing networks with application to ride-hailing. arXiv preprint arXiv

Banerjee, S., Riquelme, C., \& Johari, R. (2015). Pricing in ride-share platforms: A queueing-theoretic approach. Available at SSRN 2568258.

Barth, M., \& Boriboonsomsin, K. (2009). Traffic congestion and greenhouse gases.

Basset, P. M., Vu, B. D., Beaumier, P., Reboul, G., \& Ortun, B. (2018). Models and methods at ONERA for the presizing of EVTOL hybrid aircraft including analysis of failure scenarios. In Annual Forum Proceedings - AHS International (Vol. 2018-May). American Helicopter Society.
Battifarano, M., \& Qian, Z. S. (2019). Predicting real-time surge pricing of ride-sourcing companies. Transportation Research Part C: Emerging Technologies, 107, 444-462.

Becker, K., Terekhov, I., \& Gollnick, V. (2018). A global gravity model for air passenger demand between city pairs and future interurban air mobility markets identification. In 2018 aviation technology, integration, and operations conference (p. 2885).

Bilen, D. (2018). Introducing Ride Pass: A new way to plan your day (and budget) with Uber. Retrieved from https://www.uber.com/newsroom/ride-pass/

Bimpikis, K., Candogan, O., \& Saban, D. (2019). Spatial pricing in ridesharing networks. Operations Research, 67(3), 744-769.

Binder, R., Garrow, L. A., German, B., Mokhtarian, P., Daskilewicz, M., \& Douthat, T. H. (2018). If You Fly It, Will Commuters Come? A Survey to Model Demand for eVTOL Urban Air Trips. In 2018 Aviation Technology, Integration, and Operations Conference (p. 2882).

Board, N. S., \& National Research Council. (2005). Autonomous vehicles in support of naval operations. National Academies Press.

Boddupalli, S. S. (2019). Estimating demand for an electric vertical landing and takeoff (eVTOL) air taxi service using discrete choice modeling (Doctoral dissertation, Georgia Institute of Technology).

Bozorgi-Amiri, A., Tavakoli, S., Mirzaeipour, H., \& Rabbani, M. (2017). Integrated locating of helicopter stations and helipads for wounded transfer under demand location uncertainty. The American journal of emergency medicine, 35(3), 410-417.

Brake, J. (2008). Identifying appropriate options for delivering urban transportation to older people. Urban Transport Xiv: Urban Transport and the Environment in the 21st Century; Brebbia, CA, Ed, 57-66.

Bugaj, M., Urminský, T., Rostáš, J., \& Pecho, P. (2019). Aircraft maintenance reserves–new approach to optimization. Transportation Research Procedia, 43, 31-40.

Bye, R. J., Johnsen, S. O., \& Lillehammer, G. (2016). Different accident levels between offshore and onshore helicopter operations—a comparison of socio-technical systems. Risk, Reliability and Safety: Innovating Theory and Practice: Proceedings of ESREL 2016 (Glasgow, Scotland, 25-29 September 2016), 59.

Campbell, I., Ali, M. M., \& Silverwood, M. (2019). Solving a dial-a-flight problem using composite variables. TOP, 1-31.

CarterCopter Technology. (2013). Retrieved 9 June 2020, from \href{https://aerospaceblog.wordpress.com/2013/01/19/cartercopter-technology/}{\nolinkurl{https://aerospaceblog.wordpress.com/2013/01/19/cartercopter-technology/}}

Cervero, R., \& Griesenbeck, B. (1988). Factors influencing commuting choices in suburban labor markets: a case analysis of Pleasanton, California. Transportation Research Part A: General, 22(3), 151-161.

Chan, F., Cheung, A., Ip, W. H., Lu, D., \& Lai, C. L. (2005). An aircraft service scheduling model using genetic algorithms. Journal of Manufacturing Technology Management.

Chauhan, S. S., \& Martins, J. R. R. A. (2019). Tilt-Wing eVTOL Takeoff Trajectory Optimization. Journal of Aircraft, 1–20.

Chee, W. L., \& Fernandez, J. L. (2013). Factors that influence the choice of mode of transport in Penang: A preliminary analysis. Procedia-Social and Behavioral Sciences, 91, 120-127.

Chen, J. C., Wen, C. K., Jin, S., \& Wong, K. K. (2016). A low complexity pilot scheduling algorithm for massive MIMO. IEEE Wireless Communications Letters, 6(1), 18-21.

Chen, L., \& Wilson, C. (2017). Observing algorithmic marketplaces in-the-wild. ACM SIGecom Exchanges, 15(2), 34-39.

Chen, M. K., \& Sheldon, M. (2015). Dynamic pricing in a labor market: Surge pricing and flexible work on the Uber platform. UCLA Anderson. URL: https://www. anderson. ucla. edu.

Chen, X. M., Zahiri, M., \& Zhang, S. (2017). Understanding ridesplitting behavior of on-demand ride services: An ensemble learning approach. Transportation Research Part C: Emerging Technologies, 76, 51-70.

Choi, T. M., Wallace, S. W., \& Wang, Y. (2018). Big data analytics in operations management. Production and Operations Management, 27(10), 1868-1883.

Choi, T. M., Wen, X., Sun, X., \& Chung, S. H. (2019). The mean-variance approach for global supply chain risk analysis with air logistics in the blockchain technology era. Transportation Research Part E: Logistics and Transportation Review, 127, 178-191.

Chung, S. H., Ma, H. L., Hansen, M., \& Choi, T. M. (2020). Data science and analytics in aviation. Transportation Research Part E: Logistics and Transportation Review, 134. 

Corberán, Á., Landete, M., Peiró, J., \& Saldanha-da-Gama, F. (2020). The facility location problem with capacity transfers. Transportation Research Part E: Logistics and Transportation Review, 138, 101943.

Cordeau, J. F., Stojković, G., Soumis, F., \& Desrosiers, J. (2001). Benders decomposition for simultaneous aircraft routing and crew scheduling. Transportation science, 35(4), 375-388.

Cui, R., Dong, X., \& Lin, Y. (2019). Models for aircraft maintenance routing problem with consideration of remaining time and robustness. Computers \& Industrial Engineering, 137, 106045.

Dandl, F., Hyland, M., Bogenberger, K., \& Mahmassani, H. S. (2019). Evaluating the impact of spatio-temporal demand forecast aggregation on the operational performance of shared autonomous mobility fleets. Transportation, 46(6), 1975-1996.

Datta A. (2018). Commercial Intra City On-Demand Electric-VTOL Status of Technology. An AHS/NARI Transformative Vertical Flight Working Group -2 Report. Retrieved from: \href{https://vtol.org/files/dmfile/TVF.WG2.YR2017draft.pdf}{\nolinkurl{https://vtol.org/files/dmfile/TVF.WG2.YR2017draft.pdf}}

Davis, N., Raina, G., \& Jagannathan, K. (2016, November). A multi-level clustering approach for forecasting taxi travel demand. In 2016 IEEE 19th International Conference on Intelligent Transportation Systems (ITSC) (pp. 223-228). IEEE.

Davis, N., Raina, G., \& Jagannathan, K. (2018). Taxi demand forecasting: A HEDGE-based tessellation strategy for improved accuracy. IEEE Transactions on Intelligent Transportation Systems, 19(11), 3686-3697.

Deckert, W. H., \& Franklin, J. A. (1989). Powered-lift aircraft technology (Vol. 501). Scientific and Technical Information Branch, National Aeronautics and Space Administration.

Deng, Q., Santos, B. F., \& Curran, R. (2020). A practical dynamic programming based methodology for aircraft maintenance check scheduling optimization. European Journal of Operational Research, 281(2), 256-273.

Díaz-Ramírez, J., Huertas, J. I., \& Trigos, F. (2014). Aircraft maintenance, routing, and crew scheduling planning for airlines with a single fleet and a single maintenance and crew base. Computers \& Industrial Engineering, 75, 68-78.

Downs, A. (2005). Still stuck in traffic: coping with peak-hour traffic congestion. Brookings Institution Press.

Dunn, N. (2018). Analysis of urban air transportation operational constraints and customer value attributes (Doctoral dissertation, Massachusetts Institute of Technology).

Duval, D. T. (2013). Critical issues in air transport and tourism. Tourism Geographies, 15(3), 494-510.

El Moudani, W., \& Mora-Camino, F. (2000). A dynamic approach for aircraft assignment and maintenance scheduling by airlines. Journal of Air Transport Management, 6(4), 233-237.

Enconniere, J., Ortiz-Carretero, J., \& Pachidis, V. (2017). Mission performance analysis of a conceptual coaxial rotorcraft for air taxi applications. Aerospace Science and Technology, 69, 1-14.

Engineer, F. G., Nemhauser, G. L., \& Savelsbergh, M. W. (2011). Dynamic programming-based column generation on time-expanded networks: Application to the dial-a-flight problem. INFORMS Journal on Computing, 23(1), 105-119.

Erkan, T. E., \& Elsharida, W. M. (2019). Overview of Airport Location Selection Methods. International Journal of Applied Engineering Research, 14(7), 1613-1618.

Espinoza, D., Garcia, R., Goycoolea, M., Nemhauser, G. L., \& Savelsbergh, M. W. (2008a). Per-seat, on-demand air transportation part I: Problem description and an integer multicommodity flow model. Transportation Science, 42(3), 263-278.

Espinoza, D., Garcia, R., Goycoolea, M., Nemhauser, G. L., \& Savelsbergh, M. W. (2008b). Per-seat, on-demand air transportation part II: Parallel local search. Transportation Science, 42(3), 279-291.

Ezzinbi, O., Sarhani, M., El Afia, A., \& Benadada, Y. (2014, June). A metaheuristic approach for solving the airline maintenance routing with aircraft on ground problem. In 2014 International Conference on Logistics Operations Management (pp. 48-52). IEEE.

Farahani, R. Z., SteadieSeifi, M., \& Asgari, N. (2010). Multiple criteria facility location problems: A survey. Applied Mathematical Modelling, 34(7), 1689-1709.

Feng, G., Kong, G., \& Wang, Z. (2020). We are on the way: Analysis of on-demand ride-hailing systems. Manufacturing \& Service Operations Management.

Fernando, W. D. T. (2018, November). Tri-Engine Single Thrust Vector Analysis for Yaw Movement of Flying Wing. In 2018 IEEE 5th International Conference on Engineering Technologies and Applied Sciences (ICETAS) (pp. 1-6). IEEE.

Friedrich, M., Hartl, M., \& Magg, C. (2018). A modeling approach for matching ridesharing trips within macroscopic travel demand models. Transportation, 45(6), 1639-1653.

Furuhata, M., Dessouky, M., Ordóñez, F., Brunet, M. E., Wang, X., \& Koenig, S. (2013). Ridesharing: The state-of-the-art and future directions. Transportation Research Part B: Methodological, 57, 28-46.

Gao, Z., Lin, Z., LaClair, T. J., Liu, C., Li, J. M., Birky, A. K., \& Ward, J. (2017). Battery capacity and recharging needs for electric buses in city transit service. Energy, 122, 588-600.

Garrow, L. A., German, B., Mokhtarian, P., Daskilewicz, M., Douthat, T. H., \& Binder, R. (2018). If You Fly It, Will Commuters Come? A Survey to Model Demand for eVTOL Urban Air Trips. In 2018 Aviation Technology, Integration, and Operations Conference (p. 2882).

Garrow, L. A., German, B., Mokhtarian, P., \& Glodek, J. (2019). A Survey to Model Demand for eVTOL Urban Air Trips and Competition with Autonomous Ground Vehicles. In AIAA Aviation 2019 Forum (p. 2871).

Giannini, F., Kaufman, A., \& Kearney, M. (2018). Configuration development and subscale flight testing of an urban mobility eVTOL. In Proceedings of the AHS International Technical Meeting on Aeromechanics Design for Transformative Vertical Flight 2018. American Helicopter Society  International.

Giges N., Demand for Flying Taxis Lifts Electric Aircraft Market. (2020). Retrieved 13 February 2020, from https://www.asme.org/topics-resources/content/demand-flying-taxis-lifts-electric-aircraft

Gregory, K. B., Winn, W., Johnson, K., \& Rosekind, M. R. (2010). Pilot fatigue survey: exploring fatigue factors in air medical operations. Air medical journal, 29(6), 309-319.

Guda, H., \& Subramanian, U. (2019). Your uber is arriving: Managing on-demand workers through surge pricing, forecast communication, and worker incentives. Management Science, 65(5), 1995-2014.

Halvorson, B., Halvorson, B., Editor, B., Writer, Editor, B., \& Writer, S. et al. (2019). Electrified Bell Nexus concept shows how Uber Air could take off. Retrieved 9 June 2020, from \href{https://www.greencarreports.com/news/1121971_electrified-bell-nexus-concept-shows-how-uber-air-could-take-off}{\nolinkurl{https://www.greencarreports.com/news/1121971_electrified-bell-nexus-concept-shows-how-uber-air-could-take-off}}

Hammad, A. W., Akbarnezhad, A., \& Rey, D. (2017). Bilevel mixed-integer linear programming model for solving the single airport location problem. Journal of Computing in Civil Engineering, 31(5), 06017001.

Hammad, A. W., Akbarnezhad, A., \& Rey, D. (2017). Sustainable urban facility location: Minimising noise pollution and network congestion. Transportation research part E: logistics and transportation review, 107, 38-59.

Hasan, S. (2019). Urban Air Mobility (UAM) Market Study. Retrieved 20 February 2020, from https://ntrs.nasa.gov/search.jsp?R=20190026762

Hawkins. Hyundai will make flying cars for Uber’s air taxi service. Retrieved 1 March 2020, from https://www.theverge.com/2020/1/6/21048373/hyundai-flying-car-uber-air-taxi-ces-2020

Herbawi, W., \& Weber, M. (2012, June). The ridematching problem with time windows in dynamic ridesharing: A model and a genetic algorithm. In 2012 IEEE Congress on Evolutionary Computation (pp. 1-8). IEEE.

Hess, Daniel Baldwin. 2001. “Effect of Free Parking on Commuter Mode Choice: Evidence from Travel Diary Data.” Transportation Research Record: Journal of the Transportation Research Board 1753(1):35–42.

Ho, S. C., Szeto, W. Y., Kuo, Y. H., Leung, J. M., Petering, M., \& Tou, T. W. (2018). A survey of dial-a-ride problems: Literature review and recent developments. Transportation Research Part B: Methodological, 111, 395-421.

Hoffmann, G. M., Huang, H., Waslander, S. L., \& Tomlin, C. J. (2007). Quadrotor helicopter flight dynamics and control: Theory and experiment. In Collection of Technical Papers - AIAA Guidance, Navigation, and Control Conference 2007 (Vol. 2, pp. 1670–1689).

Holden and Goel. “Fast-Forwarding to a Future of On-Demand Urban Air Transportation.” 27 October 2016.

Holden, J., Allison, E., Goel, N., \& Swaintek, S, (2018). Session presented at the meeting of the Uber Keynote: Scaling UberAir.

Hornyak, T. The flying taxi market may be ready for takeoff, changing the travel experience forever. Retrieved 4 June 2020, from https://www.cnbc.com/2020/03/06/the-flying-taxi-market-is-ready-to-change-worldwide-travel.html

Howard C., Bell Nexus full-scale air taxi design debuts, taps novel VTOL, hybrid propulsion, energy storage, and avionics technologies. (2020). Retrieved 13 February 2020, from \href{https://www.sae.org/news/2019/01/bell-nexus-full-scale-air-taxi-design-debuts-taps-novel-vtol-hybrid-propulsion-energy-storage-and-avionics-technologies}{\nolinkurl{https://www.sae.org/news/2019/01/bell-nexus-full-scale-air-taxi-design-debuts-taps-novel-vtol-hybrid-propulsion-energy-storage-and-avionics-technologies}}

Hu, M., \& Zhou, Y. (2018). Dynamic type matching. Rotman School of Management Working Paper, (2592622).

Ikeuchi, H., Hatoyama, K., Kusakabe, R., \& Kariya, I. (2019). Development of a Statistical Model to Predict Traffic Congestion in Winter Seasons in Nagaoka, Japan Using Publicly Available Data. In Intelligent Transport Systems for Everyone’s Mobility (pp. 265-278). Springer, Singapore.

INRIX. “INRIX Global Traffic Scorecard.” INRIX - INRIX, 2018, inrix.com/scorecard/.

Irizarry, R. (2003). Restructuring the spaces under elevated expressways: a case study of the spaces below the Interstate-10 overpass at Perkins Road in Baton Rouge, Louisiana.

Jiang, S., Chen, W., Li, Z., \& Yu, H. (2019). Short-term demand prediction method for online car-hailing services based on a least squares support vector machine. IEEE Access, 7, 11882-11891.

Johnson, W., Silva, C., \& Solis, E. (2018). Concept vehicles for VTOL air taxi operations. In Proceedings of the AHS International Technical Meeting on Aeromechanics Design for Transformative Vertical Flight 2018. American Helicopter Society International.

Kawabata, M., \& Shen, Q. (2006). Job accessibility as an indicator of auto-oriented urban structure: a comparison of Boston and Los Angeles with Tokyo. Environment and Planning B: Planning and Design, 33(1), 115-130.

Ke, J., Zheng, H., Yang, H., \& Chen, X. M. (2017). Short-term forecasting of passenger demand under on-demand ride services: A spatio-temporal deep learning approach. Transportation Research Part C: Emerging Technologies, 85, 591-608.

Kienzler, M., \& Kowalkowski, C. (2017). Pricing strategy: A review of 22 years of marketing research. Journal of Business Research, 78, 101-110.

King, F. H. (1986). Aviation maintenance management. SIU Press.

Kirschstein, T. (2020). Comparison of energy demands of drone-based and ground-based parcel delivery services. Transportation Research Part D: Transport and Environment, 78, 102209.

Korolko, N., Woodard, D., Yan, C., \& Zhu, H. (2018). Dynamic pricing and matching in ride-hailing platforms. Available at SSRN.

Koźlak, A., \& Wach, D. (2018). Causes of traffic congestion in urban areas. Case of Poland. In SHS Web of Conferences (Vol. 57, p. 01019). EDP Sciences.

Kumari, S., Karn, H., Varshney, L., Garg, L., \& Kumari, N. (2020, February). Energy Management Strategy for Cost Minimization under Time of Use Pricing for Residential Application. In 2020 IEEE International Students' Conference on Electrical, Electronics and Computer Science (SCEECS) (pp. 1-4). IEEE.

Kung, L. C., \& Zhong, G. Y. (2017). The optimal pricing strategy for two-sided platform delivery in the sharing economy. Transportation Research Part E: Logistics and Transportation Review, 101, 1-12.

Lam, C. T., \& Liu, M. (2017). Demand and consumer surplus in the on-demand economy: the case of ride sharing. Social Science Electronic Publishing, 17(8), 376-388.

Lei, C., Jiang, Z., \& Ouyang, Y. (2020). Path-based dynamic pricing for vehicle allocation in ridesharing systems with fully compliant drivers. Transportation Research Part B: Methodological.

Li, M. Z., Ryerson, M. S., \& Balakrishnan, H. (2019). Topological data analysis for aviation applications. Transportation Research Part E: Logistics and Transportation Review, 128, 149-174.

Li, T., \& Wan, Y. (2019). Estimating the geographic distribution of originating air travel demand using a bi-level optimization model. Transportation Research Part E: Logistics and Transportation Review, 131, 267-291.

Lin, K., Wu, J., \& Liu, D. (2019). Economic Efficiency Analysis of Micro Energy Grid Considering Time-of-Use Gas Pricing. IEEE Access, 8, 3016-3028.


Liu, Y., Wang, T., Zhang, H., Cheutet, V., \& Shen, G. (2019). The design and simulation of an autonomous system for aircraft maintenance scheduling. Computers \& Industrial Engineering, 137, 106041.

Lu, Z., Liu, Y., Debiasi, M., \& Khoo, B. C. (2016). Acoustic characteristics of a multi-rotor MAV and its noise reduction technology. In Proceedings of the INTER-NOISE 2016 - 45th International Congress and Exposition on Noise Control Engineering: Towards a Quieter Future (pp. 725–735). German Acoustical Society (DEGA).

Luo, H., Cai, J., Zhang, K., Xie, R., \& Zheng, L. (2020). A multi-task deep learning model for short-term taxi demand forecasting considering spatiotemporal dependences. Journal of Traffic and Transportation Engineering (English Edition).

Luo, Q., \& Saigal, R. (2017). Dynamic pricing for on-demand ridesharing: A continuous approach. Available at SSRN 3056498.

Ma, H., Fang, F., \& Parkes, D. C. (2018). Spatio-temporal pricing for ridesharing platforms. arXiv preprint arXiv:1801.04015.

Ma, S., Zheng, Y., \& Wolfson, O. (2014). Real-time city-scale taxi ridesharing. IEEE Transactions on Knowledge and Data Engineering, 27(7), 1782-1795.

Ma, T. Y., Rasulkhani, S., Chow, J. Y., \& Klein, S. (2019). A dynamic ridesharing dispatch and idle vehicle repositioning strategy with integrated transit transfers. Transportation Research Part E: Logistics and Transportation Review, 128, 417-442.

Hamandi, M., Usai, F., Sablé, Q., Staub, N., Tognon, M., \& Franchi, A. (2020). Survey on Aerial Multirotor Design: a Taxonomy Based on Input Allocation.

Ma, T. (2017). Passenger transport systems based on pilotless vertical takeoff and landing (vtol) aircraft. U.S. Patent Application 15/256,754.

Masson, R., Lehuédé, F., \& Péton, O. (2014). The dial-a-ride problem with transfers. Computers \& Operations Research, 41, 12-23.

Matyas, M., \& Kamargianni, M. (2019). Survey design for exploring demand for Mobility as a Service plans. Transportation, 46(5), 1525-1558.

Mercier, A., Cordeau, J. F., \& Soumis, F. (2005). A computational study of Benders decomposition for the integrated aircraft routing and crew scheduling problem. Computers \& Operations Research, 32(6), 1451-1476.

Moore J. (2020). Kitty Hawk's self-flying taxi takes flight. Retrieved 13 February, 2020, from \href{https://www.aopa.org/news-and-media/all-news/2018/march/14/self-flying-taxi-takes-flight}{\nolinkurl{https://www.aopa.org/news-and-media/all-news/2018/march/14/self-flying-taxi-takes-flight}}

Moreira-Matias, L., Gama, J., Ferreira, M., Mendes-Moreira, J., \& Damas, L. (2013). Predicting taxi–passenger demand using streaming data. IEEE Transactions on Intelligent Transportation Systems, 14(3), 1393-1402.

Mueller, E. R., Kopardekar, P. H., \& Goodrich, K. H. (2017). Enabling airspace integration for high-density on-demand mobility operations. In 17th AIAA Aviation Technology, Integration, and Operations Conference (p. 3086).

Munari, P. (2017). Mathematical modeling in the airline industry: optimizing aircraft assignment for on-demand air transport. Proceeding Series of the Brazilian Society of Computational and Applied Mathematics, 5(1).

Najmi, A., Rey, D., \& Rashidi, T. H. (2017). Novel dynamic formulations for real-time ride-sharing systems. Transportation research part E: logistics and transportation review, 108, 122-140.

Nelson, J. D., Wright, S., Masson, B., Ambrosino, G., \& Naniopoulos, A. (2010). Recent developments in flexible transport services. Research in Transportation Economics, 29(1), 243-248.

Nelson, T., \& Orton, F. (2013). A new approach to congestion pricing in electricity markets: Improving user pays pricing incentives. Energy Economics, 40, 1-7.

Ng, W., \& Datta, A. (2019). Hydrogen Fuel Cells and Batteries for Electric-Vertical Takeoff and Landing Aircraft. Journal of Aircraft, 56(5), 1765-1782.

O’Hear, (2017). Lilium, a German company building an electric ‘air taxi,’ makes key hires from Gett, Airbus and Tesla. Retrieved 21 February, 2020, from \href{https://techcrunch.com/2017/08/22/lilium/}{\nolinkurl{https://techcrunch.com/2017/08/22/lilium/}}

\"Ozkan, E., \& Ward, A. R. (2020). Dynamic matching for real-time ride sharing. Stochastic Systems, 10(1), 29-70.

Papakostas, N., Papachatzakis, P., Xanthakis, V., Mourtzis, D., \& Chryssolouris, G. (2010). An approach to operational aircraft maintenance planning. Decision Support Systems, 48(4), 604-612.

Partnership for New York City. “Growth or Gridlock: The Economic Case for Traffic Relief and Transit Improvement for a Greater New York.” Partnership for New York City, Partnership for New York City, 2019, \href{www.pfnyc.org/reports/GrowthGridlock-4pg.pdf}{\nolinkurl{www.pfnyc.org/reports/GrowthGridlock-4pg.pdf}}.

Patnoe, L. (2018). Flyshare 2020 (Doctoral dissertation). 

Percoco, M. (2010). Airport activity and local development: evidence from Italy. Urban studies, 47(11), 2427-2443.


Polaczyk, N., Trombino, E., Wei, P., \& Mitici, M. (2015). A review of current technology and research in urban on-demand air mobility applications. In 8th Biennial Autonomous VTOL Technical Meeting and 6th Annual Electric VTOL Symposium 2019 (pp. 333–343). Vertical Flight Society.

Polaczyk, N., Trombino, E., Wei, P., \& Mitici, M. (2019). A review of current technology and research in urban on-demand air mobility applications. In 8th Biennial Autonomous VTOL Technical Meeting and 6th Annual Electric VTOL Symposium.

Pradeep P. and Wei P. (August, 2018). Energy optimal speed profile for arrival of tandem tilt-wing e-VTOL aircraft with RTA constraint. IEEE/CSAA Guidance, Navigation and Control Conference (GNCC), Xiamen, China.

Pradeep, P., \& Wei, P. (2019). Energy-Efficient Arrival with RTA Constraint for Multirotor eVTOL in Urban Air Mobility. Journal of Aerospace Information Systems, 16(7), 263-277.

Qi, X., Bard, J. F., \& Yu, G. (2004). Class scheduling for pilot training. Operations Research, 52(1), 148-162.

Qian, X., Zhang, W., Ukkusuri, S. V., \& Yang, C. (2017). Optimal assignment and incentive design in the taxi group ride problem. Transportation Research Part B: Methodological, 103, 208-226.

Qin, Y., Chan, F. T., Chung, S. H., \& Qu, T. (2017, April). Development of MILP model for integrated aircraft maintenance scheduling and multi-period parking layout planning problems. In 2017 4th International Conference on Industrial Engineering and Applications (ICIEA) (pp. 197-203). IEEE.

Qiu, H., Li, R., \& Zhao, J. (2018). Dynamic pricing in shared mobility on demand service. arXiv preprint arXiv:1802.03559.


Rajendran, S, \& Pagel, E. (2020). Recommendations for Emerging Air Taxi Network Operations based on Online Review Analysis of Helicopter Services. arXiv preprint arXiv: 2006.10898.

Rajendran, S \& Shulman, J. (2020). Study of Emerging Air Taxi Network Operation using Discrete-Event Systems Simulation Approach. Journal of Air Transport Management (in press).

Rajendran, S., \& Zack, J. (2019). Insights on strategic air taxi network infrastructure locations using an iterative constrained clustering approach. Transportation Research Part E: Logistics and Transportation Review, 128, 470-505.

Rath, S., \& Chow, J. Y. (2019). Air Taxi Skyport Location Problem for Airport Access. arXiv preprint arXiv:1904.01497.

Reddy, D. T. (2018). Agent based simulation of the dial-a-flight problem (Doctoral dissertation).

Reiche, C., McGillen, C., Siegel, J., \& Brody, F. (2019, April). Are We Ready to Weather Urban Air Mobility (UAM)?. In 2019 Integrated Communications, Navigation and Surveillance Conference (ICNS) (pp. 1-7). IEEE.

Rodrigues, F., Markou, I., \& Pereira, F. C. (2019). Combining time-series and textual data for taxi demand prediction in event areas: A deep learning approach. Information Fusion, 49, 120-129.

Ross, P. E. (2018). The prius of the sky [Blueprints for a Miracle]. IEEE Spectrum, 55(6).

Rothfeld, R., Balac, M., Ploetner, K. O., \& Antoniou, C. (2018). Initial Analysis of Urban Air Mobility’s Transport Performance in Sioux Falls. In 2018 Aviation Technology, Integration, and Operations Conference (p. 2886).

Ryll, M., Bülthoff, H. H., \& Giordano, P. R. (2012). Modeling and control of a quadrotor UAV with tilting propellers. In Proceedings - IEEE International Conference on Robotics and Automation (pp. 4606–4613)

Saharan, S., Bawa, S., \& Kumar, N. (2020). Dynamic pricing techniques for Intelligent Transportation System in smart cities: A systematic review. Computer Communications, 150, 603-625.

Salama, M., \& Srinivas, S. (2020). Joint optimization of customer location clustering and drone-based routing for last-mile deliveries. Transportation Research Part C: Emerging Technologies, 114, 620-642.

Sampson, B. Rolls-Royce reveals 300mph electric aircraft. Retrieved 9 June 2020, from \href{https://www.aerospacetestinginternational.com/news/electric-hybrid/rolls-royce-reveals-300mph-electric-aircraft.html}{\nolinkurl{https://www.aerospacetestinginternational.com/news/electric-hybrid/rolls-royce-reveals-300mph-electric-aircraft.html}}

Sanchez, D. T., Boyacı, B., \& Zografos, K. G. (2020). An optimisation framework for airline fleet maintenance scheduling with tail assignment considerations. Transportation Research Part B: Methodological, 133, 142-164. 

Santi, P., Resta, G., Szell, M., Sobolevsky, S., Strogatz, S. H., \& Ratti, C. (2014). Quantifying the benefits of vehicle pooling with shareability networks. Proceedings of the National Academy of Sciences, 111(37), 13290-13294.

Sanyé-Mengual, E., Anguelovski, I., Oliver-Solà, J., Montero, J. I., \& Rieradevall, J. (2016). Resolving differing stakeholder perceptions of urban rooftop farming in Mediterranean cities: promoting food production as a driver for innovative forms of urban agriculture. Agriculture and human values, 33(1), 101-120.

Saurin, T. A., \& Junior, G. C. C. (2012). A framework for identifying and analyzing sources of resilience and brittleness: a case study of two air taxi carriers. International Journal of Industrial Ergonomics, 42(3), 312-324.

Schwanen, T., \& Mokhtarian, P. L. (2005). What affects commute mode choice: neighborhood physical structure or preferences toward neighborhoods?. Journal of transport geography, 13(1), 83-99.

Shahriari, M., Bozorgi-Amiri, A., Tavakoli, S., \& Yousefi-Babadi, A. (2017). Bi-objective approach for placing ground and air ambulance base and helipad locations in order to optimize EMS response. The American journal of emergency medicine, 35(12), 1873-1881.

Shamiyeh, M., Rothfeld, R., \& Hornung, M. (2018). A performance benchmark of recent personal air vehicle concepts for urban air mobility. In 31st Congress of the International Council of the Aeronautical Sciences, ICAS 2018. International Council of the Aeronautical Sciences.

Shen, Q., Chen, P., \& Pan, H. (2016). Factors affecting car ownership and mode choice in rail transit-supported suburbs of a large Chinese city. Transportation Research Part A: Policy and Practice, 94, 31-44.


Shladover, S. E. (2018). Connected and automated vehicle systems: Introduction and overview. Journal of Intelligent Transportation Systems, 22(3), 190-200.

Silva, C., Johnson, W., Antcliff, K. R., \& Patterson, M. D. (2018). VTOL urban air mobility concept vehicles for technology development. In 2018 Aviation Technology, Integration, and Operations Conference. American Institute of Aeronautics and Astronautics Inc, AIAA.

Sivasubramaniyam, R. D., Charlton, S. G., \& Sargisson, R. J. (2020). Mode choice and mode commitment in commuters. Travel Behaviour and Society, 19, 20-32.

Smith, R. D. (1994). Safe Heliports Through Design and Planning. A Summary of FAA Research and Development (No. DOT/FAA/RD-93/17). Federal Aviation Administration Washington DC Systems Research and Development Service.

Smith, R. D. (2001). Heliport/Vertiport Design Deliberations 1997-2000 (No. DOT/FAA/ND-00/1). Federal Aviation Administration Washington DC Associate Administrator for NAS Development.

Specht, K., \& Sanyé-Mengual, E. (2017). Risks in urban rooftop agriculture: Assessing stakeholders’ perceptions to ensure efficient policymaking. Environmental science \& policy, 69, 13-21.

Sriram, C., \& Haghani, A. (2003). An optimization model for aircraft maintenance scheduling and re-assignment. Transportation Research Part A: Policy and Practice, 37(1), 29-48.

Staplin, L., Mastromatto, T., Lococo, K. H., Kenneth W. Gish, K. W., \& Brooks, J. O. (2018, September). The effects of medical conditions on driving performance (Report No. DOT HS 812 623). Washington, DC: National Highway Traffic Safety Administration

Stojković, M., \& Soumis, F. (2001). An optimization model for the simultaneous operational flight and pilot scheduling problem. Management Science, 47(9), 1290-1305.

Stopka, U., Pessier, R., \& Günther, C. (2018, July). Mobility as a service (MaaS) based on intermodal electronic platforms in public transport. In International Conference on Human-Computer Interaction (pp. 419-439). Springer, Cham.

Straubinger, A., Rothfeld, R., Shamiyeh, M.,…, \& Plötner, K. (2020). An overview of current research and developments in urban air mobility – Setting the scene for UAM introduction. Journal of Air Transport Management (in press).

Suh, D. Y., \& Ryerson, M. S. (2019). Forecast to grow: Aviation demand forecasting in an era of demand uncertainty and optimism bias. Transportation Research Part E: Logistics and Transportation Review, 128, 400-416.

Sun, L., Teunter, R. H., Babai, M. Z., \& Hua, G. (2019). Optimal pricing for ride-sourcing platforms. European Journal of Operational Research, 278(3), 783-795.

Sun, X., Wandelt, S., \& Stumpf, E. (2018). Competitiveness of on-demand air taxis regarding door-to-door travel time: A race through Europe. Transportation Research Part E: Logistics and Transportation Review, 119, 1-18.

Swadesir, L., \& Bil, C. (2019). Urban Air Transportation for Melbourne Metropolitan Area. In AIAA Aviation 2019 Forum (p. 3572).

Tarafdar, S., Rimjha, M., Hinze, N., Hotle, S., \& Trani, A. A. (2019, April). Urban air Mobility Regional Landing Site Feasibility and Fare Model Analysis in the Greater Northern California Region. In 2019 Integrated Communications, Navigation and Surveillance Conference (ICNS) (pp. 1-11). IEEE.

The Vertical Flight Society (2020). Retrieved 9 May 2020, from https://evtol.news/aircraft/

Thomaier, S., Specht, K., Henckel, D., Dierich, A., Siebert, R., Freisinger, U. B., \& Sawicka, M. (2015). Farming in and on urban buildings: Present practice and specific novelties of Zero-Acreage Farming (ZFarming). Renewable Agriculture and Food Systems, 30(1), 43-54.

Tyrinopoulos, Y., \& Antoniou, C. (2013). Factors affecting modal choice in urban mobility. European Transport Research Review, 5(1), 27-39.

Van der Zwan, F. M., Wils, K., \& Ghijs, S. S. A. (2011). Development of an aircraft routing system for an air taxi operator. In Aeronautics and Astronautics. IntechOpen.

Vascik, P. D., \& Hansman, R. J. (2018). Scaling Constraints for Urban Air Mobility Operations: Air Traffic Control, Ground Infrastructure, and Noise. In 2018 Aviation Technology, Integration, and Operations Conference (p. 3849).

Ventura Diaz, P., Johnson, W., Ahmad, J., \& Yoon, S. (2019). The Side-by-Side Urban Air Taxi Concept. In AIAA Aviation 2019 Forum (p. 2828).

Wang, J., Lim, M. K., Zhan, Y., \& Wang, X. (2020). An intelligent logistics service system for enhancing dispatching operations in an IoT environment. Transportation Research Part E: Logistics and Transportation Review, 135, 101886.

Wang, X., He, F., Yang, H., \& Gao, H. O. (2016). Pricing strategies for a taxi-hailing platform. Transportation Research Part E: Logistics and Transportation Review, 93, 212-231.

Wang, Y., \& Li, L. (2015). Time-of-use electricity pricing for industrial customers: A survey of US utilities. Applied Energy, 149, 89-103.

Wang, Y., Winter, S., \& Ronald, N. (2017). How much is trust: The cost and benefit of ridesharing with friends. Computers, Environment and Urban Systems, 65, 103-112.

Watanabe, C., Naveed, K., \& Neittaanmäki, P. (2016). Co-evolution of three mega-trends nurtures un-captured GDP–Uber’s ride-sharing revolution. Technology in Society, 46, 164-185.

Wheeler, K. (2019). Lyft announces membership plan for \$19.99 a month: Lyft Pink. Retrieved June 3, 2020, from \href{https://www.usatoday.com/story/money/2019/10/29/lyft-membership-plan-lyft-pink/2484960001/}{\nolinkurl{https://www.usatoday.com/story/money/2019/10/29/lyft-membership-plan-lyft-pink/2484960001/}}

Xavier, I. R., de Mello Bandeira, R. A., de Oliveira Silva, L., Bandeira, A. D. P. F., \& Campos, V. B. G. (2019). Employing helicopters in the modelling of last mile distribution system in large-scale disasters. Transportation research procedia, 37, 306-313.

Xu, J., Rahmatizadeh, R., Bölöni, L., \& Turgut, D. (2017). Real-time prediction of taxi demand using recurrent neural networks. IEEE Transactions on Intelligent Transportation Systems, 19(8), 2572-2581.

Yan, C., Zhu, H., Korolko, N., \& Woodard, D. (2019). Dynamic pricing and matching in ride‐hailing platforms. Naval Research Logistics (NRL).

Yang, H., Fung, C. S., Wong, K. I., \& Wong, S. C. (2010). Nonlinear pricing of taxi services. Transportation Research Part A: Policy and Practice, 44(5), 337-348.

Yang, Z., Yu, S., \& Notteboom, T. (2016). Airport location in multiple airport regions (MARs): The role of land and airside accessibility. Journal of Transport Geography, 52, 98-110.

Ye, R., \& Titheridge, H. (2017). Satisfaction with the commute: The role of travel mode choice, built environment and attitudes. Transportation Research Part D: Transport and Environment, 52, 535-547.

Yu, B., Guo, Z., Asian, S., Wang, H., \& Chen, G. (2019). Flight delay prediction for commercial air transport: A deep learning approach. Transportation Research Part E: Logistics and Transportation Review, 125, 203-221.

Yu, G., Dugan, S., \& Argüello, M. (1998). Moving toward an integrated decision support system for manpower planning at Continental Airlines: Optimization of pilot training assignments. In Industrial Applications of Combinatorial Optimization (pp. 1-24). Springer, Boston, MA.

Yun, L., Wang, X., Fan, H., \& Li, X. (2020). Reliable facility location design with round-trip transportation under imperfect information Part I: A discrete model. Transportation Research Part E: Logistics and Transportation Review, 133, 101825.

Zha, L., Yin, Y., \& Du, Y. (2017). Surge Pricing and Labor Supply in the Ride-Sourcing Market. Transportation Research Procedia, 23, 2-21.

Zha, L., Yin, Y., \& Xu, Z. (2018). Geometric matching and spatial pricing in ride-sourcing markets. Transportation Research Part C: Emerging Technologies, 92, 58-75.

Zhang, M., Zhang, Y., Qiu, Z., \& Wu, H. (2019). Two-Stage Covering Location Model for Air–Ground Medical Rescue System. Sustainability, 11(12), 3242.

Zhang, W., Shemshadi, A., Sheng, Q. Z., Qin, Y. L., Xu, X., \& Yang, J. (2018). A User-Oriented Taxi Ridesharing System with Large-Scale Urban GPS Sensor Data. IEEE Transactions on Big Data.

Zhao, K., Khryashchev, D., \& Vo, H. (2019). Predicting Taxi and Uber Demand in Cities: Approaching the Limit of Predictability. IEEE Transactions on Knowledge and Data Engineering.

Zhou, J. (2012). Sustainable commute in a car-dominant city: Factors affecting alternative mode choices among university students. Transportation research part A: policy and practice, 46(7), 1013-1029.

\end{hangparas}

\end{document}